\documentclass[preprintnumbers,amsmath,amssymb,twocolumn,superscriptaddress]{revtex4-2}

\usepackage{mathrsfs}
\usepackage{amssymb}
\usepackage{graphicx}
\usepackage{dcolumn}
\usepackage{bm}
\usepackage{textcomp}
\usepackage{amsmath}
\usepackage[titletoc]{appendix}
\usepackage{accents}
\usepackage{ntheorem}
\usepackage{color}
\usepackage{datetime} 

\newcommand\ket[1]{\ensuremath{|#1\rangle}}

\newcommand\iprod[2]{\ensuremath{\langle#1|#2\rangle}}
\newcommand\oprod[2]{\ensuremath{|#1\rangle\langle#2|}}
\newcommand\mean[1]{\ensuremath{\langle #1\rangle}}

\newcounter{RomanNumber}

\makeatletter
\def\widebar{\accentset{{\cc@style\underline{\mskip10mu}}}}
\def\Widebar{\accentset{{\cc@style\underline{\mskip8mu}}}}
\makeatother

\begin{document}

\title{The practical issues of side-channel-secure quantum key distribution}
\author{Cong Jiang}\email{Corresponding author: jiangcong@jiqt.org}
\affiliation{Jinan Institute of Quantum Technology and Jinan branch, Hefei National Laboratory, Jinan, Shandong 250101, China}
\affiliation{State Key Laboratory of Low Dimensional Quantum Physics, Department of Physics, Tsinghua University, and Frontier Science Center for Quantum Information, Beijing 100084, China}

\author{Xiao-Long Hu}
\affiliation{School of Physics and Optoelectronic Engineering, Guangdong University of Technology, Guangzhou, 510006, China}

\author{Zong-Wen Yu}
\affiliation{Data Communication Science and Technology Research Institute, Beijing 100191, China}

\author{Hai Xu}
\affiliation{State Key Laboratory of Low Dimensional Quantum Physics, Department of Physics, Tsinghua University, and Frontier Science Center for Quantum Information, Beijing 100084, China}

\author{Xiang-Bin Wang}\email{Corresponding author: xbwang@mail.tsinghua.edu.cn}
\affiliation{Jinan Institute of Quantum Technology and Jinan branch, Hefei National Laboratory, Jinan, Shandong 250101, China}
\affiliation{State Key Laboratory of Low Dimensional Quantum Physics, Department of Physics, Tsinghua University, and Frontier Science Center for Quantum Information, Beijing 100084, China}
\affiliation{ International Quantum Academy, Shenzhen 518048, China.}

\begin{abstract}
Quantum Key Distribution (QKD) leverages the principles of quantum mechanics to provide theoretically unconditional security for cryptographic key sharing. However, practical implementations remain vulnerable due to non-ideal devices and potential security loopholes at both the source and detection sides of QKD systems. The side-channel-secure (SCS) protocol addresses these challenges by encoding bits in vacuum and non-vacuum states and introducing a third-party measurement node, thereby repelling attacks targeting the detection side as well as external lab attacks on the source side. In this work, we consider the state-dependent correlated errors and Trojan-horse attack while preserving the SCS protocol's key advantage—specifically, requiring only upper bounds on intensities characterization without needing a full description of quantum states in infinite dimensions. Numerical results demonstrate that when the reflected light intensity from Trojan-horse attacks falls below $10^{-6}$, Eve can scarcely extract additional key information from the reflections. This work makes the SCS protocol more practical.
\end{abstract}



\maketitle
\section{Introduction}\label{intro}
Quantum key distribution (QKD) exploits the laws of quantum mechanics to guarantee unconditional security in cryptographic key exchange, making it a leading candidate for future secure communications~\cite{bennett1984quantum,gisin2002quantum,xu2020secure,pirandola2020advances,scarani2009security}. Since the introduction of the BB84 protocol in 1984~\cite{bennett1984quantum}, substantial theoretical and experimental progress has dramatically enhanced QKD’s security, key generation rates, and achievable transmission distances~\cite{liu2023experimental,pittaluga2025long, pittaluga2021600, chen2021twin, liu2021field, wang2022twin, zhou2023twin, li2023high}. Nevertheless, practical implementations remain susceptible to device imperfections, environmental disturbances, and various side channels, leaving a gap between idealized security proofs and real-world deployments.

To eliminate detector side vulnerabilities, measurement-device-independent (MDI) QKD~\cite{lo2012measurement,braunstein2012side,wang2013three,zhou2016making,jiang2021higher} and twin-field QKD protocols~\cite{lu2018overcoming,wang2018twin,ma2018phase,lin2018simple,curty2018simple,cui2019twin} were developed, completely immune to attacks targeting the detection side. On the source side, the decoy-state method mitigates risks from imperfect single-photon sources~\cite{hwang2003quantum,wang2005beating,lo2005decoy}. Yet infinite dimensional side channels—such as variations in waveform, frequency~\cite{huang2018quantum}, and time-varying encodings~\cite{gnanapandithan2025hidden}—still threaten security, due to the infeasibility of fully characterizing quantum states. The side-channel-secure (SCS) QKD protocol was proposed to address these issues by only bounding the intensity of emitted signals rather than requiring a full infinite-dimensional description~\cite{wang2019practical}. A recent experimental demonstration over a 50 km fiber link has confirmed its practical viability~\cite{zhang2022experimental}.

In our recent works, we have introduced an SCS protocol utilizing non-ideal vacuum sources and proven its security equivalence to the original SCS scheme~\cite{jiang2023side}. We also derived a finite-key analysis via a mapping approach to calculate secure key rates (in this work, the key rate means \textbf{key rate per pulse}) under realistic block sizes~\cite{jiang2024side}. However, the security of the current SCS protocol still relies on two fundamental assumptions:
\\1. The set of candidate states for all time windows is the direct product of the candidate states for each time window.
\\2. Eve has no access to inside Alice’s (Bob’s) lab. More precisely, we request that Eve has no access to anything that stores Alice's (Bob's) secret information inside lab and also the initial source states inside lab are independent of Eve's probe.  

As shown in Sec.~\ref{review}, Assumption 1 actually requires that the state in the $j$-th time window is independent of the state selection information of other time windows. If the practical system cannot maintain this condition, then the state independence requirement for each time window in the SCS protocol is not satisfied, and therefore its security needs to be reconsidered. Assumption 2 prohibits any intrusion into the lab, yet both Alice's and Bob's sources emit quantum states outward, leaving them vulnerable to Trojan-horse attacks~\cite{xu2020secure, gisin2006trojan, lucamarini2015practical, tan2021chip}. While optical isolators and one-way components can mitigate such threats, they cannot entirely nullify their influence on the key rate. Thus, a comprehensive security proof must explicitly incorporate the Troja-horse attack.

In this paper, we eliminate both assumptions while preserving the principal advantage of SCS‑QKD: only upper bounds on the intensities must be characterized, without requiring a full infinite‑dimensional state description. This advancement bridges the gap between theoretical security and experimental implementation.

\section{Review of the SCS protocol}\label{review}
In the previous SCS protocol~\cite{wang2019practical,jiang2023side,jiang2024side}, there are two sources named "$o$" and "$x$" which produce candidate states on Alice's side (Bob's side). For clarity, we shall focus our analysis on Alice's operations, with the understanding that parallel arguments hold for Bob's implementation. The protocol's robustness is ensured through intensity differentiation between sources $o$ and $x$. In general, source $o$ is a source whose intensity is as close to $0$ as possible, that is, a vacuum source or a non-ideal vacuum source. Notably, the security proof of the SCS protocol primarily requires establishing upper bounds for the source intensities rather than complete characterization of the infinite dimensions space.

In order to better distinguish from the logical window defined in Sec.~\ref{corre}, we rename the time window defined in the previous SCS protocol to the physical window. Within each physical window, Alice probabilistically commits to classical bit values $0$ or $1$ with respective probabilities $p_0$ and $p_x=1-p_0$. For bit $0$, Alice prepares a candidate state from the source $o$; for bit $1$, Alice prepares a candidate state from the source $x$. Assumption 1 in Sec.~\ref{intro} actually requires that the state in the $j$-th time window is independent of the state selection information of other time windows. For simplicity, we will refer to this requirement as the \textbf{independence condition} later. In what follows, we would prove that the independence condition leads to the Assumption 1.

Without loss of generality, the states from sources $o$ and $x$ can be written into the form of $\ket{\psi_{{o_j}|\bar{\mathcal{L}}_j}^{A}}_{C_j}$ and $\ket{\psi_{{x_j}|\bar{\mathcal{L}}_j}^{A}}_{C_j}$ respectively, where $\bar{\mathcal{L}}_j$ contains the state choices (i.e., the bit values) of all other time windows. Then the states prepared by Alice in the whole protocol can be written as
\begin{equation}\label{rhoA}
\rho_A=\sum_{r_1,\cdots, r_N}\bigotimes_{j=1}^N p_{r_j}\oprod{r}{r}_{A_j} \otimes \oprod{\psi_{{r_j}|\bar{\mathcal{L}}_j}^{A}}{\psi_{{r_j}|\bar{\mathcal{L}}_j}^{A}}_{C_j},
\end{equation} 
where $r_j=o,x$ for all $j$, $A_j$ is the local memories that stores the state choice of the $j$-th physical window, $C_j$ is the subsystem that sent to Charlie in the $j$-th physical window, and $N$ is the total number of physical windows in the protocol. According to the definition of $\bar{\mathcal{L}}_j$, we have 
\begin{equation}
\bar{\mathcal{L}}_j=\{r_1,\cdots, r_{j-1},r_{j+1},\cdots, r_N\}.
\end{equation}

The independence condition can be transformed into the following equation
\begin{equation}\label{condt}
\ket{\psi_{{r_j}|\bar{\mathcal{L}}_j}^{A}}_{C_j}=\ket{\psi_{{r_j}|\bar{\mathcal{L}}_{j}^\prime}^{A}}_{C_j},
\end{equation}
for all $\bar{\mathcal{L}}_j\neq \bar{\mathcal{L}}_{j}^\prime$. Thus we can denote
\begin{equation}
\ket{\psi_{{r_j}}^{A}}_{C_j} \equiv \ket{\psi_{{r_j}|\bar{\mathcal{L}}_j}^{A}}_{C_j}.
\end{equation}

Then, we can rewrite Eq.~\eqref{rhoA} as 
\begin{equation}\label{rhoA2}
\begin{split}
\rho_A=&\sum_{r_1,\cdots, r_N}\bigotimes_{j=1}^N p_{r_j}\oprod{r}{r}_{A_j} \otimes \oprod{\psi_{{r_j}}^{A}}{\psi_{{r_j}}^{A}}_{C_j}\\
=&\bigotimes_{j=1}^N \sum_{r_j}p_{r_j}\oprod{r}{r}_{A_j} \otimes \oprod{\psi_{{r_j}}^{A}}{\psi_{{r_j}}^{A}}_{C_j}\\
=&\bigotimes_{j=1}^N (p_o\oprod{o_j}{o_j}_{A_j} \otimes \ket{\psi_{{o_j}}^{A}}_{C_j}\\
&+p_x\oprod{x_j}{x_j}_{A_j} \otimes \ket{\psi_{{x_j}}^{A}}_{C_j}). 
\end{split}
\end{equation} 

The above process proves that if the state in the $j$-th time window is independent of the state selection information of other time windows, the set of candidate states for all time windows is the direct product of the candidate states for each time window. With this, we can prove the security of the real protocol by proving that the candidate states of every physical window in an ideal protocol can be mapped to the candidate states in the real protocol through virtual attenuation and unitary transformations~\cite{jiang2024side}. 

Note that the in the security proof of Ref.~\cite{jiang2024side}, we only require the independence condition, but the states in different physical windows can exhibit correlations. For example, QKD systems can have the following properties:  the intensities of all states in the first $N/2$ physical windows are relatively higher $10\%$ than their default values, and those of the second $N/2$ physical windows are relative weaker $10\%$ than their default values. In this case, the states of different physical windows are correlated, but the independence condition still holds. 

\section{Security proof for SCS protocol with correlated errors}\label{corre}
\subsection{The main idea}
In the practical QKD systems, especially the high speed systems, the source may exhibits state-dependent correlated errors: the state in the $j$-th physical window could depend on the source settings of the preceding $\xi$ windows~\cite{roberts2018patterning, zapatero2021security, pereira2024quantum}. Mathematically, this results in Eq.~\eqref{condt} no longer holding. And therefore its security needs to be reconsidered. In this work, we consider the case of $\xi=1$, i.e., the case where correlation exists only between adjacent pulses. However, our results can be easily extended to case $\xi > 1$.

As shown in Sec.~\ref{review}, if the independence condition holds, i.e., there is no state-dependent correlated errors, the states prepared by Alice and Bob can be written into the following form~\cite{jiang2024side}
\begin{equation}\label{real_statep2}
\begin{split}
&\ket{\psi_{x _j}^{A}}=\sqrt {a_{0j}}  \ket{0}+\sqrt{1- a_{0j} } \ket{\tilde \psi_{x _j}^{A}},\\
&\ket{\psi_{x _j}^{B}}=\sqrt {b_{0j}} \ket{0}+\sqrt {1- b_{0j}} \ket{\tilde \psi_{x _j}^{B}},
\end{split}
\end{equation}
for the candidate states from the source $x$ of Alice and Bob respectively. Here $\ket{0}$ is the vacuum state and $\ket{\tilde \psi_{x _j}^{A}}$ ($\ket{\tilde \psi_{x _j}^{B}}$) is a whole-space state containing at least one photon. And
\begin{equation}\label{real_state1}
\begin{split}
&\ket{\psi_{o _j}^{A}}=\sqrt{a_{v0j}}\ket{0}+\sqrt{1-a_{v0j}}\ket{\tilde \psi_{o _j}^{A}},\\
&\ket{\psi_{o _j}^{B}}=\sqrt{b_{v0j}}\ket{0}+\sqrt{1-b_{v0j}}\ket{\tilde\psi_{o _j}^{B}},
\end{split}
\end{equation}
for the candidate states from source $o$ of Alice and Bob respectively. Here state $\ket{\tilde \psi_{o _j}^{A}}$ ($\ket{\tilde\psi_{o _j}^{B}}$) is a whole-space states containing at least one photon.

If, however, the source exhibits state-dependent correlated errors, the states prepared by Alice and Bob in each physical window will take the form of $\ket{\psi_{r_j|t}^{S}}$ for $r,t=o,x$ and $S=A,B$, and $\ket{\psi_{r_j|o}^{S}}\neq \ket{\psi_{r_j|x}^{S}}$. Here, the subscript $r_j|t$ indicates the preparation of an state $r_j$ in the $j$-th physical windows under the condition that the state in the preceding physical window was $t$. This means we can no longer direct apply the conclusion of  Ref.~\cite{jiang2024side}. To address this issue, in this work, we define logical windows which contains several physical windows. In the previous SCS protocol, the bits are encoded into different states of the physical window. But in this work, the bits are encoded into different states of the logical window. Since we primarily consider $\xi=1$, we define each logical window to contain two physical windows, with the state of the second physical window being vacuum to achieve the highest key rate. For bit value 1 (0) in the $i$-th logical window, Alice (Bob) actually prepares the following state:
\begin{equation}\label{real_statep2c}
\begin{split}
&\ket{\phi_{w _i}^{A}}=\ket{\psi_{x _i|o}^{A}}\otimes \ket{\psi_{o _i|x}^{A}},\\
&\ket{\phi_{w _i}^{B}}=\ket{\psi_{x _i|o}^{B}}\otimes \ket{\psi_{o _i|x}^{B}}.
\end{split}
\end{equation}
where
\begin{equation}\label{corre_state}
\begin{split}
&\ket{\psi_{x_i|o}^{A}}=\sqrt {a_{0i|o}}  \ket{0}+\sqrt{1- a_{0i|o} } \ket{\tilde \psi_{x _i|o}^{A}},\\
&\ket{\psi_{x_i|o}^{B}}=\sqrt {b_{0i|o}} \ket{0}+\sqrt {1- b_{0i|o}} \ket{\tilde \psi_{x _i|o}^{B}},\\
&\ket{\psi_{o_i|x}^{A}}=\sqrt{a_{v0i|x}}\ket{0}+\sqrt{1-a_{v0i|x}}\ket{\tilde \psi_{o _i|x}^{A}},\\
&\ket{\psi_{o_i|x}^{B}}=\sqrt{b_{v0i|x}}\ket{0}+\sqrt{1-b_{v0i|x}}\ket{\tilde\psi_{o _i|x}^{B}}.
\end{split}
\end{equation}

Note that in Eq.~\eqref{real_statep2c}, $\ket{\phi_{w_i}^{A}}$ is the state of the $i$-th logical window which contains two states from the ($j=2i-1$)-th and ($j^\prime=2i$)-th physical windows, $\ket{\psi_{x _j|o}^{A}}$ and  $\ket{\psi_{o _{j^\prime}|x}^{A}}$. Without causing confusion, we simply refer to state $\ket{\psi_{x _j|o}^{A}}$ and  $\ket{\psi_{o _{j^\prime}|x}^{A}}$ as $\ket{\psi_{x _i|o}^{A}}$ and $\ket{\psi_{o_i|x}^{A}}$ respectively.

Alice (Bob) actually prepares the following state for bit value 0 (1)
\begin{equation}\label{real_statep1c}
\begin{split}
&\ket{\phi_{v _i}^{A}}=\ket{\psi_{o _i|o}^{A}}\otimes \ket{\psi_{o _i|o}^{A}},\\
&\ket{\phi_{v _i}^{B}}=\ket{\psi_{o _i|o}^{B}}\otimes \ket{\psi_{o _i|o}^{B}},
\end{split}
\end{equation}
where the definitions of $\ket{\psi_{o _i|o}^{A}}$ and $\ket{\psi_{o _i|o}^{B}}$ are similar to those of Eq.~\eqref{corre_state}.

\subsection{The protocol}
Similarly to Ref.~\cite{jiang2024side}, the protocol here contains the following main steps.

To clarify, we regard the states $\ket{\phi_{w _i}^{A}}$ or $\ket{\phi_{w _i}^{B}}$ are from the source $w$ and the states $\ket{\phi_{v _i}^{A}}$ or $\ket{\phi_{v _i}^{B}}$ are from the source $v$.

\textbf{Step 1.} In the \textit{i}-th logical window, Alice (Bob) randomly prepare a pulse (which actually contains two sub-pulses) from source $w$ or $v$ with probabilities $p_w$ and $p_{v}=1-p_{w}$. These pulses are called signal pulses. Alice (Bob) also prepares a strong reference pulse time-multiplexed with each of the signal pulses. 

\textbf{Step 2.} After receiving the signal pulses, Charlie first performs phase compensation and then interferometry measurement at his measurement station. The measurement results would be announced to Alice and Bob. If the right-side detector clicks, it is regarded as an effective logical window.

Remark: In this work, we assume that when two identical coherent lights arrive at Charlie's detection station, most of the energy reaches the left-side detector. We also define $\mathcal{O}$ ($\mathcal{B}$) windows as logical windows when Alice and Bob both decide on source $v$ ($w$), and define $\mathcal{Z}$ windows as logical windows when only one of Alice or Bob decides on source $v$, the bits of $\mathcal{Z}$ windows are untagged bits~\cite{jiang2024side}.

\textbf{Step 3.} After preparing $N$ logical windows and received all measurement results, Alice and Bob use data from effective logical windows for the raw bits. In the data postprocessing, they first perform error correction. After this, Alice and Bob shall know the values of $n_{\zeta }$, ($\zeta$ = $\mathcal{O}$, $\mathcal{B}$, $\mathcal{Z}$), where $n_{\zeta}$ is the number of effective $\zeta$ windows.

We assume the following conditions hold for all logical windows $i$ with $t=x,o$
\begin{equation}\label{boundA}
\begin{split}
    &a_{v0i|t} \ge  {a_{v0}} \ge 0.5;\;
    a_{0i|t} \ge  {a_{0}}\ge 0.5,\\
     &b_{v0i|t} \ge  {b_{v0}} \ge 0.5;\;
    b_{0i|t} \ge  {b_{0}}\ge 0.5,
    \end{split}
\end{equation}
where ${a_{v0}},{a_{0}},{b_{v0}},{b_{0}}$ are the lower bounds of their corresponding physical quantities respectively.

\subsection{The security proof}
Although the physical windows are correlated with each other, Eqs.~(\ref{real_statep2c},\ref{real_statep1c}) show that the states of different logical windows satisfy the independence condition. Therefore, we can use the security proof method proposed in Ref.~\cite{jiang2024side}. 

The fidelity of the two states in the $i$-th logical window in Alice's side is
\begin{equation}
f_{Ai} = |\iprod{\phi_{w_i}^A}{\phi_{v_i}^A}|^2. 
\end{equation}
Given the bounds of Eq.~\eqref{boundA}, for any $i$, we obtain the following range $S_{RA}$ for fidelity  $f_{Ai}\in S_{RA}$:
\begin{equation}\label{attr}
  S_{RA} = \left[ \left( G(a_0,a_{v0}) \cdot G(a_{v0},a_{v0}) \right)^2,\, 1 \right]
\end{equation}
where
\begin{equation}
G(\alpha,\beta)=\sqrt{ \alpha \beta}-\sqrt{(1-{\alpha})(1-\beta)}.
\end{equation}

Similarly, given the bounds of Eq.~\eqref{boundA}, we can obtain the following range $S_{RB}$ for fidelity  $f_{Bi}\in S_{RB}$:
\begin{equation}\label{attrb}
S_{RB} = \left[\left[G(b_0,b_{v0})\cdot G(b_{v0},{b_{v0}})\right]^2, 1\right],
\end{equation}
where
\begin{equation}
f_{Bi} = |\iprod{\phi_{w_i}^B}{\phi_{v_i}^B}|^2. 
\end{equation}

As shown in Ref.~\cite{jiang2024side}, if there exists a map $\mathcal{M}$ that maps the states in a virtual perfect protocol to those in the real protocol through attenuation and unitary transformation, we can prove the security of the real protocol through proving the security of the virtual perfect protocol.

For ease of presentation, we assume the states in each logical window of Alice (Bob) in the virtual perfect protocol are perfect weak coherent states and perfect vacuum states
\begin{equation}\label{virtual_statep1}
\begin{split}
&\ket{\mu_{v}^A}=\ket{0},\ket{\mu_{w}^A} = \sum_{k=0}^\infty \frac{e^{-\mu_{A}/2}\mu_{A}^{k/2}}{\sqrt k!}\ket{k},\\
&\ket{\mu_{v}^B}=\ket{0},\ket{\mu_{w}^B} = \sum_{k=0}^\infty \frac{e^{-\mu_{B}/2}\mu_{B}^{k/2}}{\sqrt k!}\ket{k}.
\end{split}
\end{equation}

The fidelity $F_A$ ($F_B$) of Alice's (Bob's) two states in the virtual perfect protocol is
\begin{equation}
\begin{split}
&F_A=\left| \iprod{\mu_{v}^A}{\mu_{w}^A}\right|^2=e^{-\mu_A},\\
&(F_B=\left| \iprod{\mu_{v}^B}{\mu_{w}^B}\right|^2=e^{-\mu_B}.)
\end{split}
\end{equation}

If the following conditions are satisfied
\begin{equation}\label{equiva1}
\begin{split}
&e^{-\mu_A}=\left[G(a_v,a_{v0})\cdot G(a_{v0}{a_{v0}})\right]^2,\\
&e^{-\mu_B}=\left[G(b_v,b_{v0})\cdot G(b_{v0},{b_{v0}})\right]^2,
\end{split}
\end{equation}
the map $\mathcal{M}$ exists~\cite{jiang2024side}. Thus we can prove the security of the real protocol through proving the security of the virtual perfect protocol. Since the two conditioned states in the virtual perfect protocol are the same for all time windows, we can use the post-selection technique to prove the protocol's security against coherent attacks.

\subsection{The key rate formulas with finite-key effect}
With Eq.~\eqref{equiva1}, we can use the key rate formulas for the perfect protocol to calculate the secure key rate of the real protocol~\cite{wang2019practical,jiang2024side}. 

With the values of $n_{\zeta}$, the upper bound of expected value of the number of phase-flip errors $\mean{\bar{N}^{ph}}$ can be calculated by 
\begin{equation}
\begin{split}
&\mean{\bar{N}^{ph}}=\frac{p_vp_w}{2}\left[\frac{\mean{n_{\mathcal{O}}}^U}{p_v^2}+\frac{\mean{n_{\mathcal{B}}}^U}{p_w^2}+\frac{2\bar{c}_2}{p_v}\sqrt{N\mean{n_{\mathcal{O}}}^U}\right.\\
&+ \frac{2\bar{c}_2}{p_w}\sqrt{N\mean{n_{\mathcal{B}}}^U}+\frac{2}{p_vp_w}\sqrt{\mean{n_{\mathcal{O}}}^U\mean{n_{\mathcal{B}}}^U}+\left.\bar{c}_2^2N \right],
\end{split}
\end{equation}
where 
\begin{equation}\label{c2}
   \bar{c}_2=2\sqrt{(1-e^{-\mu_A/2})(1-e^{-\mu_B/2})}.
\end{equation}
Here $\mean{\cdot}$ is the expected value of a physical quantity, and the superscript $U$ represents the upper bound of the expected value under a certain failure probability obtained when estimating the expected value from its corresponding observation. In this work, we apply the Chernoff bound to perform this estimation~\cite{chernoff1952measure}.

With Chernoff bound~\cite{chernoff1952measure}, we also have the upper bound of real value of the number of phase-flip errors $\bar{N}^{ph}$
\begin{equation}
\bar{N}^{ph}=O^U(\mean{\bar{N}^{ph}}),
\end{equation}  
where $O^U(\cdot)$ is defined in Eq.~\eqref{OU}. The upper bound of the real value of the phase-flip error rate is
\begin{equation}
\bar{e}^{ph}=\bar{N}^{ph}/n_{\mathcal{Z}}.
\end{equation}

Then, Alice and Bob can calculate the secure final key rate under collective attack by~\cite{scarani2008security,sheridan2010finite}
\begin{equation}\label{keyrate}
\begin{split}
&R_{\text{col}}=\frac{1}{N}\left\{n_{\mathcal{Z}}[1-H(\bar{e}^{ph})]-\text{leak}_{\text{EC}}-\log_2\frac{2}{\varepsilon_{cor}}\right .\\
&\left .-2\log_2\frac{1}{\varepsilon_{PA}}-(d+3)\sqrt{n_{\mathcal{Z}}\log_2\frac{2}{\bar{\varepsilon}}} \right\}.
\end{split}
\end{equation}
Here, $\text{leak}_{\text{EC}}$ is the amount of information leakage during the error correction process and generally $leak_{EC}=fM_sH(E_{\mathcal{Z}})$ where $f$ is the error correction inefficiency, and $M_s=n_{\mathcal{O}}+n_{\mathcal{B}}+n_{\mathcal{Z}}$ is the number of raw keys, and $E_{\mathcal{Z}}$ is the bit-flip error rate of the raw key strings; $H(x)=-x\log_2x-(1-x)\log_2(1-x)$ is the Shannon entropy; $\varepsilon_{cor}$ is the failure probability of the error correction; $\varepsilon_{PA}$ is the failure probability of the privacy amplification; $\bar{\varepsilon}$ is the coefficient of measuring the accuracy of estimating the smooth min-entropy; $d$ is the dimension of the local states shared by Alice and Bob, and $d=8$ in SCS protocol. 

While Alice and Bob perform the privacy amplification process according to Eq.~\eqref{keyrate}, the protocol is $\varepsilon_{col}$-secure under collective attack \cite{sheridan2010finite}, and
\begin{equation}
\varepsilon_{\text{col}}=\varepsilon_{cor}+\bar{\varepsilon}+\varepsilon_{PA}+3\epsilon,
\end{equation}
where $\epsilon$ is the probability which is used in the estimation of Chernoff bound.

Applying the post-selection technique~\cite{christandl2009postselection}, by shorten $2(d^2-1)\log_2(N+1)$ bits of the final key distilled under collective attack, the protocol obtains $\varepsilon_{\text{coh}}$-secure final keys against coherent attack by 
\begin{equation}\label{rcoh1}
R_{\text{coh}}=R_{\text{col}}-\frac{2(d^2-1)\log_2 (N+1)}{N},
\end{equation} 
and the corresponding security coefficient under coherent attack is~\cite{christandl2009postselection} 
\begin{equation}\label{coh}
\varepsilon_{\text{coh}}=\varepsilon_{\text{col}}(N+1)^{d^2-1}.
\end{equation}

\section{SCS protocol against Trojan-horse attack}
The Trojan-horse attack is an active attack strategy~\cite{xu2020secure}. Its basic principle is that Eve injects additional, unauthorized probe light into the QKD system during its operation. This probe light may cause internal components of the system—such as phase modulators, intensity modulators, or encoding devices—to produce reflections, scatterings, or back-reflected signals. These reflected signals may carry information about the encoding, and by measuring them, Eve could potentially obtain extra information about the transmitted quantum states or the encoding settings. Trojan-horse attacks can be launched against both the source side and the detector side~\cite{gisin2006trojan,jain2014trojan,jain2014risk}. However, since the SCS protocol is measurement-device-independent, we do not need to consider Trojan-horse attacks targeting the detector side.

To defend against Trojan-horse attacks targeting the source side, the source side typically adopts measures such as installing optical isolators, unidirectional fiber devices, or power-limiting components~\cite{tan2021chip,ponosova2022protecting}, which can significantly attenuate the incoming external light, thereby reducing the effectiveness of Trojan-horse attacks. The power of the light injected by the eavesdropper is limited by factors such as the fiber damage threshold or the power limiter, implying a maximum allowable input power. By increasing the reverse isolation at the source side, the intensity of the reflected lights from the eavesdropper can be attenuated to an extremely low level. As shown in Ref.~\cite{tan2021chip}, when the isolation reaches 230 dB, the intensity of the reflected signal drops below $10^{-18}$ per pulse.

The Trojan-horse attack can also be modeled equivalently as Alice and Bob directly preparing the following state and sending it through the channel controlled by Eve
\begin{equation}\label{th_state}
\begin{split}
&\ket{\phi_{x _i}^{\prime A}}=\ket{\psi_{x _i|o}^{A}}\otimes \ket{\psi_{o _i|x}^{A}}\otimes \ket{\psi_{e_i|w}^{A}},\\
&\ket{\phi_{x _i}^{\prime B}}=\ket{\psi_{x _i|o}^{B}}\otimes \ket{\psi_{o _i|x}^{B}}\otimes \ket{\psi_{e_i|w}^{B}},\\
&\ket{\phi_{o _i}^{\prime A}}=\ket{\psi_{o _i|o}^{A}}\otimes \ket{\psi_{o _i|o}^{A}}\otimes \ket{\psi_{e_i|v}^{A}},\\
&\ket{\phi_{o _i}^{\prime B}}=\ket{\psi_{o _i|o}^{B}}\otimes \ket{\psi_{o _i|o}^{B}}\otimes \ket{\psi_{e_i|v}^{B}},
\end{split}
\end{equation}
where 
\begin{equation}
\ket{\psi_{e_i|q}^{S}}=\sqrt{1-\mu_{e_i|q}^S}\ket{0}+\sqrt{\mu_{e_i|q}^S}\ket{\tilde\psi_{e_i|q}^{S}},
\end{equation}
and
\begin{equation}
\mu_{e_i|q}^S\le \mu_E,
\end{equation}
for $q=w,v$ and $S=A,B$. Here state $\ket{\tilde\psi_{e_i|q}^{S}}$ is a whole-space states containing at least one photon and $\mu_E$ is the upper bound of the intensity of Eve's reflected light in each logical window.

In this case, the fidelity of the two states in Alice's (Bob's) side for each logical windows is in the range $S_{RA}^\prime$ ($S_{RB}^\prime$), and  
\begin{equation}\label{attr_th}
\begin{split}
  &S_{RA}^\prime = \left[\left[G(a_0,a_{v0})\cdot G(a_{v0},{a_{v0}})\cdot G(1-\mu_E,1-\mu_E)\right]^2, 1\right],\\
   &S_{RB}^\prime = \left[\left[G(b_0,b_{v0})\cdot G(b_{v0},{b_{v0}})\cdot G(1-\mu_E,1-\mu_E)\right]^2, 1\right].
\end{split}
\end{equation}

Let
\begin{equation}\label{equiva}
\begin{split}
&e^{-\mu_A^\prime}=\left[G(a_v,a_{v0})\cdot G(a_{v0}{a_{v0}})\cdot G(1-\mu_E,1-\mu_E)\right]^2,\\
&e^{-\mu_B^\prime}=\left[G(b_v,b_{v0})\cdot G(b_{v0},{b_{v0}})\cdot G(1-\mu_E,1-\mu_E)\right]^2,
\end{split}
\end{equation}
and replace $\mu_A,\mu_B$ by $\mu_A^\prime,\mu_B^\prime$ in Eq.~\eqref{c2}, we can get the secure key rate formula for the SCS protocol against Trojan-horse attack.

\section{Numerical simulation}

The observed values are simulated by the linear model. The experimental parameters are listed in Table~\ref{exproperty}.

\begin{table}[htbp]
\centering
\begin{tabular}{ccccccc}
\hline
$p_d$& $e_d$ &$\eta_d$ & $f$ & $\alpha_f $ & $\xi_{coh}$ &$N$\\
\hline
$1.0\times10^{-9}$& {$3\%$}  & {$60.0\%$} & $1.16$ & $0.2$ & $10^{-10}$& $10^{14}$\\ 
\hline
\end{tabular}
\caption{Experimental parameters for numerical simulations. Here $p_d$ is the dark counting rate per pulse of Charlie's detectors; $e_d$ is the misalignment-error probability; $\eta_d$ is the detection efficiency of Charlie's detectors; $f$ is the error correction inefficiency; $\alpha_f$ is the fiber loss coefficient (dB/km).}\label{exproperty}
\end{table}

\begin{figure}
\centering
\includegraphics[width=9cm]{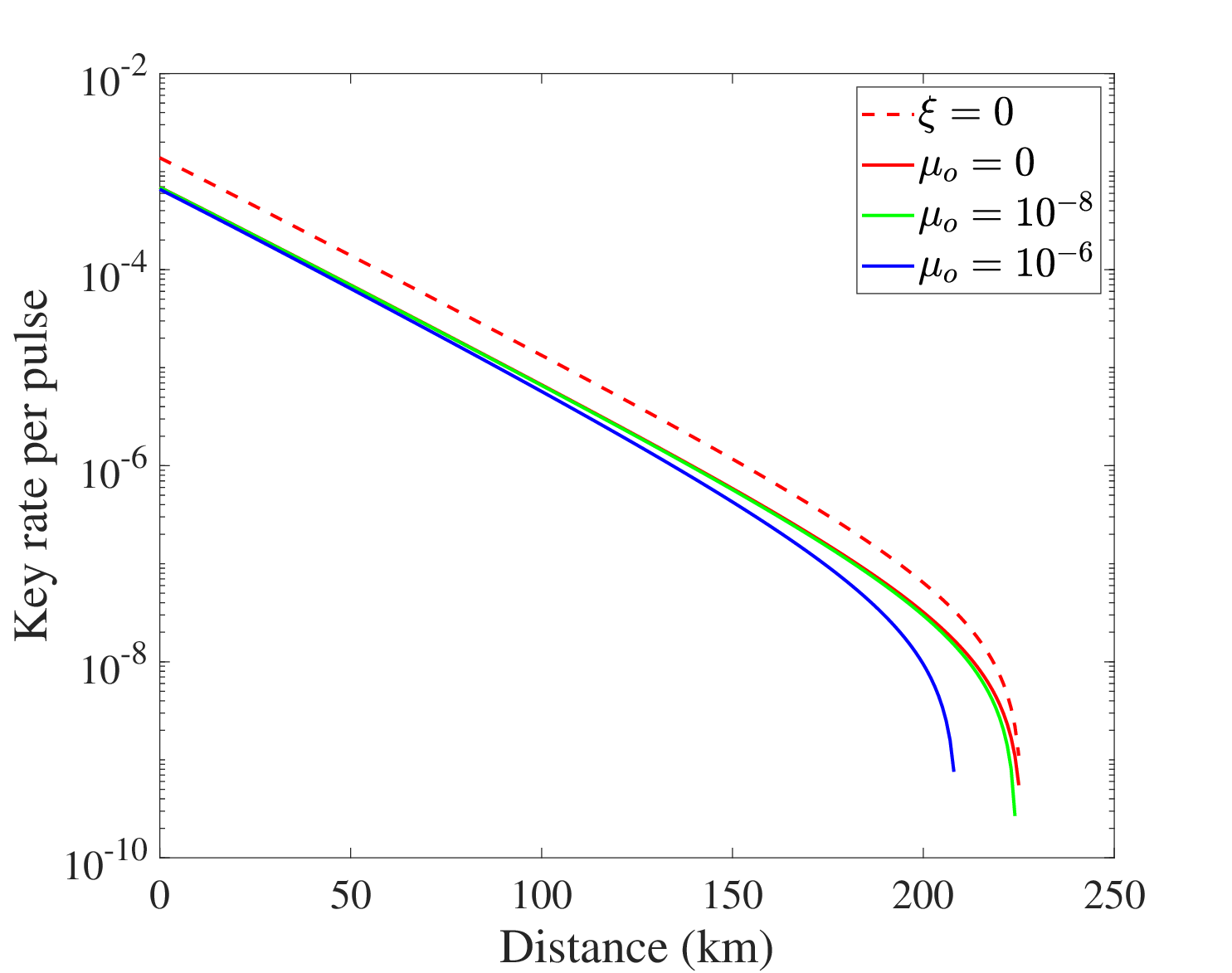}
\caption{Comparison of key rates with and without correlated errors. Different $\mu_o$ means different intensities of vacuum states.}
\label{core}
\end{figure}

Figure~\ref{core} is the comparison of the key rate with or without correlated errors where the case without correlated errors means $\xi=0$. To ensure a fair comparison with the scenario of $\xi=0$, the key rate in this section is defined as the key rate per physical window. That is, the key rate discussed in this section is half of the key rate calculated using Eq.~\eqref{rcoh1}. In the key rate calculation for the scenario of $\xi=0$, we assume a perfect vacuum. In our simulation, we consider a symmetric scenario, and let $a_0 = b_0 = e^{-\mu}$, where $\mu$ is optimized to maximize the key rate. At the same time, we define $a_{v0} = b_{v0} = e^{-\mu_o}$, which reflects the extinction ratio (ER) between the weak coherent state and the vacuum state in the system. In SCS experiments, the ER can typically be controlled to levels of 60 dB or even higher. This corresponds to $\mu_o \leq 10^{-8}$. The results shown in Figure\ref{core} indicate that when $\mu_o \leq 10^{-8}$, its impact on the key rate is negligible. Even if the ER is poorly controlled, for example $\mu_o = 10^{-6}$, the key rate shows almost no reduction at short distances, while the maximum secure transmission distance is reduced by approximately 15 km. Compared to the  scenario of $\xi=0$, the drop in key rate in the correlated error case primarily originates from the reduction in the number of effective windows. Therefore, when the ER is well-controlled, i.e., $\mu_o \leq 10^{-8}$, the key rate under correlated errors is half of that under uncorrelated errors, and the maximum achievable distance remains nearly unchanged.

\begin{figure}
\centering
\includegraphics[width=9cm]{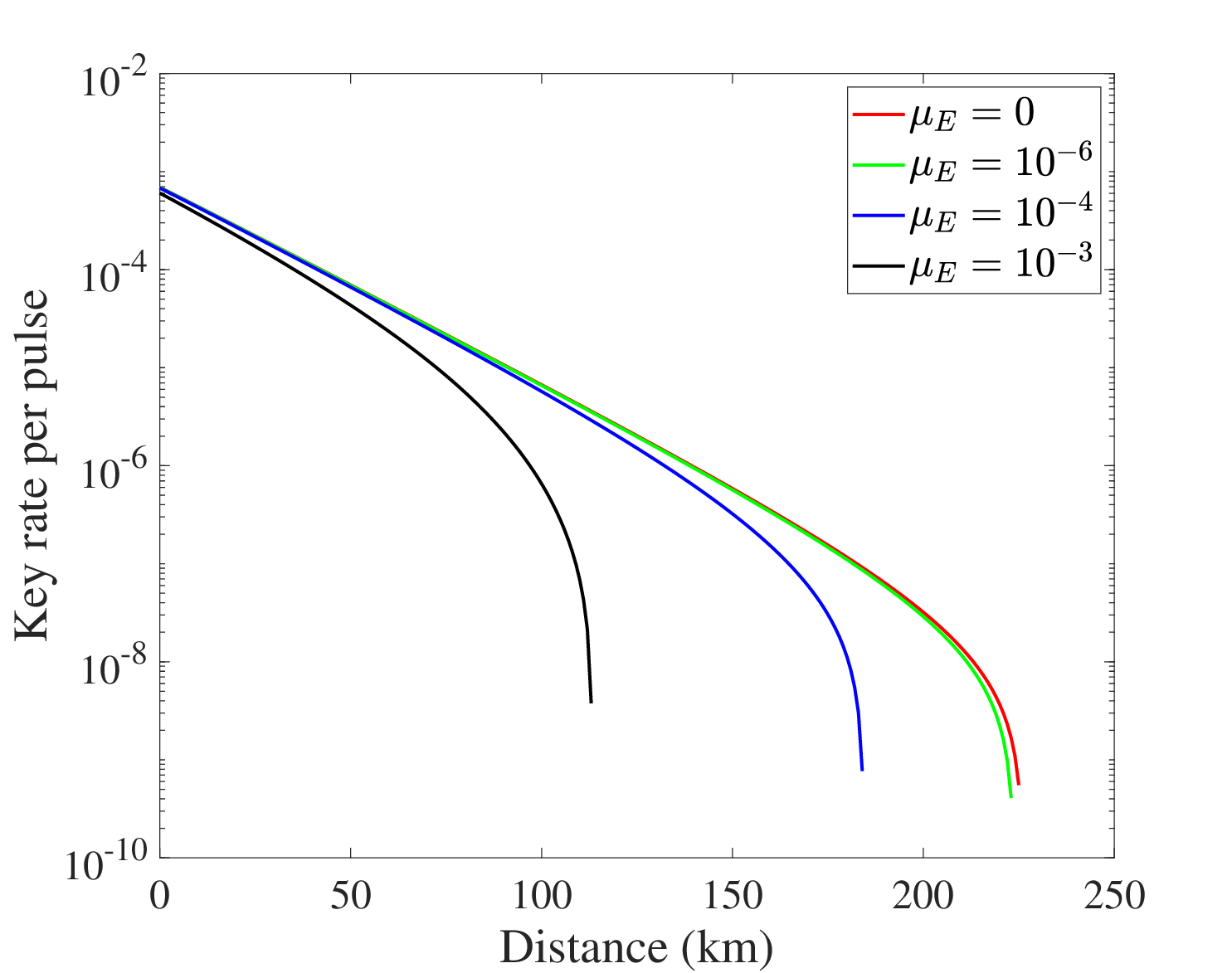}
\caption{The key rate under Trojan-horse attack. The red dashed line is the key rate without Trojan-horse attack, and we also assume a perfect vacuum in this case. For other cases, we set $\mu_o=10^{-8}$.}\label{th}
\end{figure}

Figure~\ref{th} shows the key rate under Trojan-horse attacks. The red solid line is the key rate without Trojan-horse attack, and we also assume a perfect vacuum in this case which matches the red solid line in Figure~\ref{core}. For other cases, we set $\mu_o=10^{-8}$. The results in Figure~\ref{th} show that as long as the intensity of the reflected light of the Trojan-horse attack is kept below the level of $ 10^{-6} $, the Trojan-horse attack has almost no impact on the secure key rate of the SCS protocol. Furthermore, when the intensity of the reflected light in the Trojan-horse attack is less than $10^{-4}$, the maximum secure communication distance is reduced by about 40 km. By comparing the results in Figure~\ref{core} and Figure~\ref{th}, it can be observed that in Figure~\ref{core}, when $\mu_o = 10^{-6} $, the key rate decreases significantly at long distances. However, in Figure~\ref{th}, when $\mu_E = 10^{-6}$, the key rate exhibits negligible degradation. This discrepancy arises because, although the impacts of the imperfect vacuum and the reflected light of the Trojan-horse attack on the quantum state, as shown in Eqs.~(\ref{real_statep2c},\ref{real_statep1c}) and Eq.~\eqref{th_state}, are similar, the reflected light of the Trojan-horse attack does not affect the observable values. Therefore, its influence on the key rate is relatively smaller.

\section{Discussion and conclusion}
By defining a logical window that contains multiple physical windows and encoding bits on the logical window, we demonstrate the security of the SCS protocol against correlated errors. Using a similar idea, we also incorporate Trojan horse attacks targeting the source side into our security analysis framework and present a method for calculating the secure key rate in the presence of such attacks. Numerical simulation results show that when the intensity of the non-ideal vacuum is less than $10^{-8}$, it has almost no impact on the secure key rate. When the intensity of the reflected light from a Trojan horse attack is less than $10^{-6}$,the eavesdropper gains no significant additional information from the reflected light.

Although in the main text we primarily consider the case of $\xi = 1$, i.e., the case in which only the pulses in the adjacent physical windows exhibit correlated errors, our method can be easily generalized to cases of $\xi > 1$: by defining a logical window that contains $\xi+1$ physical windows, where the latter $\xi$ physical windows emit vacuum states.

Moreover, we also consider an alternative approach based on post-selection to address the security of the SCS protocol in the presence of correlated errors. Alice and Bob follow the standard SCS protocol described in Ref.~\cite{jiang2024side}. After transmitting $N$ pulses and Charlie announces all the measurement results, Alice and Bob first reveal their state choices for the non-click events. Then, Alice and Bob perform the error correction procedure, which enables them to fully determine the distribution of windows in which both parties sent vacuum states. When $\xi = 1$, Alice and Bob first divide the transmitted pulses into odd-indexed groups, i.e., $j = 2k - 1$, and even-indexed groups, i.e., $j = 2k$ for $k = 1, 2, 3, \cdots$. An untagged bit corresponds to a window in which one party (Alice or Bob) chooses to send while the other chooses not to send, and in the subsequent window both Alice and Bob choose not to send. Then, Alice and Bob separately compute the phase-flip error rates for the odd and even groups and perform a unified privacy amplification step. Our results indicate that this method yields a slightly better key rate at short distances compared to the method discussed in the main text, but at long distances, both the key rate and the maximum achievable distance are inferior to those of the method presented in the main text.

\textbf{Funding.} This work was supported by Key R$\&$D Plan of Shandong Province Grant No. 2021ZDPT01; National Natural Science Foundation of China Grant Nos. 12374473, 12174215, 12104184; Innovation Program for Quantum Science and Technology No. 2021ZD0300705; Shandong Provincial Natural Science Foundation Grant No. ZR2021LLZ007; the Taishan Scholars Program; Leading Talents of Quancheng Industry.

\appendix
\section{Chernoff bound}\label{chernoff}
The Chernoff bound can help us estimate the expected value from their observed values~\cite{chernoff1952measure}. Let $X_1,X_2,\dots,X_n$ be $n$ independent random samples, detected with the value 1 or 0, and let $X$ denote their sum satisfying $X=\sum_{i=1}^nX_i$. $E$ is the expected value of $X$. We have
\begin{align}
\label{EL}E^L(X)=&\frac{X}{1+\delta_1(X)},\\
\label{EU}E^U(X)=&\frac{X}{1-\delta_2(X)},
\end{align}
where we can obtain the values of $\delta_1(X)$ and $\delta_2(X)$ by solving the following equations
\begin{align}
\label{delta1}\left(\frac{e^{\delta_1}}{(1+\delta_1)^{1+\delta_1}}\right)^{\frac{X}{1+\delta_1}}&=\xi,\\
\label{delta2}\left(\frac{e^{-\delta_2}}{(1-\delta_2)^{1-\delta_2}}\right)^{\frac{X}{1-\delta_2}}&=\xi,
\end{align}
where $\xi$ is the failure probability.

Besides, we can use the Chernoff bound to help us estimate their real values from their expected values. Similar to Eqs.~\eqref{EL}- \eqref{delta2}, the observed value, $O$, and its expected value, $Y$, satisfy 
\begin{align}
\label{OU}&O^U(Y)=[1+\delta_1^\prime(Y)]Y,\\
\label{OL}&O^L(Y)=[1-\delta_2^\prime(Y)]Y,
\end{align}   
where we can obtain the values of $\delta_1^\prime(Y,\xi)$ and $\delta_2^\prime(Y,\xi)$ by solving the following equations
\begin{align}
\left(\frac{e^{\delta_1^\prime}}{(1+\delta_1^\prime)^{1+\delta_1^\prime}}\right)^{Y}&=\xi,\\
\label{endd}\left(\frac{e^{-\delta_2^\prime}}{(1-\delta_2^\prime)^{1-\delta_2^\prime}}\right)^{Y}&=\xi.
\end{align}

\bibliography{refs-jiang.bib}

\begin{thebibliography}{46}%
\makeatletter
\providecommand \@ifxundefined [1]{%
 \@ifx{#1\undefined}
}%
\providecommand \@ifnum [1]{%
 \ifnum #1\expandafter \@firstoftwo
 \else \expandafter \@secondoftwo
 \fi
}%
\providecommand \@ifx [1]{%
 \ifx #1\expandafter \@firstoftwo
 \else \expandafter \@secondoftwo
 \fi
}%
\providecommand \natexlab [1]{#1}%
\providecommand \enquote  [1]{``#1''}%
\providecommand \bibnamefont  [1]{#1}%
\providecommand \bibfnamefont [1]{#1}%
\providecommand \citenamefont [1]{#1}%
\providecommand \href@noop [0]{\@secondoftwo}%
\providecommand \href [0]{\begingroup \@sanitize@url \@href}%
\providecommand \@href[1]{\@@startlink{#1}\@@href}%
\providecommand \@@href[1]{\endgroup#1\@@endlink}%
\providecommand \@sanitize@url [0]{\catcode `\\12\catcode `\$12\catcode
  `\&12\catcode `\#12\catcode `\^12\catcode `\_12\catcode `\%12\relax}%
\providecommand \@@startlink[1]{}%
\providecommand \@@endlink[0]{}%
\providecommand \url  [0]{\begingroup\@sanitize@url \@url }%
\providecommand \@url [1]{\endgroup\@href {#1}{\urlprefix }}%
\providecommand \urlprefix  [0]{URL }%
\providecommand \Eprint [0]{\href }%
\providecommand \doibase [0]{https://doi.org/}%
\providecommand \selectlanguage [0]{\@gobble}%
\providecommand \bibinfo  [0]{\@secondoftwo}%
\providecommand \bibfield  [0]{\@secondoftwo}%
\providecommand \translation [1]{[#1]}%
\providecommand \BibitemOpen [0]{}%
\providecommand \bibitemStop [0]{}%
\providecommand \bibitemNoStop [0]{.\EOS\space}%
\providecommand \EOS [0]{\spacefactor3000\relax}%
\providecommand \BibitemShut  [1]{\csname bibitem#1\endcsname}%
\let\auto@bib@innerbib\@empty
\bibitem [{\citenamefont {Bennett}\ and\ \citenamefont
  {Brassard}(1984)}]{bennett1984quantum}%
  \BibitemOpen
  \bibfield  {author} {\bibinfo {author} {\bibfnamefont {C.~H.}\ \bibnamefont
  {Bennett}}\ and\ \bibinfo {author} {\bibfnamefont {G.}~\bibnamefont
  {Brassard}},\ }\bibfield  {title} {\bibinfo {title} {Quantum cryptography:
  Public key distribution and coin tossing},\ }in\ \href@noop {} {\emph
  {\bibinfo {booktitle} {Proceedings of the IEEE International\ Conference on
  Computers, Systems, and Signal Processing}}}\ (\bibinfo {year} {1984})\ pp.\
  \bibinfo {pages} {175--179}\BibitemShut {NoStop}%
\bibitem [{\citenamefont {Gisin}\ \emph {et~al.}(2002)\citenamefont {Gisin},
  \citenamefont {Ribordy}, \citenamefont {Tittel},\ and\ \citenamefont
  {Zbinden}}]{gisin2002quantum}%
  \BibitemOpen
  \bibfield  {author} {\bibinfo {author} {\bibfnamefont {N.}~\bibnamefont
  {Gisin}}, \bibinfo {author} {\bibfnamefont {G.}~\bibnamefont {Ribordy}},
  \bibinfo {author} {\bibfnamefont {W.}~\bibnamefont {Tittel}},\ and\ \bibinfo
  {author} {\bibfnamefont {H.}~\bibnamefont {Zbinden}},\ }\bibfield  {title}
  {\bibinfo {title} {Quantum cryptography},\ }\href@noop {} {\bibfield
  {journal} {\bibinfo  {journal} {Reviews of Modern Physics}\ }\textbf
  {\bibinfo {volume} {74}},\ \bibinfo {pages} {145} (\bibinfo {year}
  {2002})}\BibitemShut {NoStop}%
\bibitem [{\citenamefont {Xu}\ \emph {et~al.}(2020)\citenamefont {Xu},
  \citenamefont {Ma}, \citenamefont {Zhang}, \citenamefont {Lo},\ and\
  \citenamefont {Pan}}]{xu2020secure}%
  \BibitemOpen
  \bibfield  {author} {\bibinfo {author} {\bibfnamefont {F.}~\bibnamefont
  {Xu}}, \bibinfo {author} {\bibfnamefont {X.}~\bibnamefont {Ma}}, \bibinfo
  {author} {\bibfnamefont {Q.}~\bibnamefont {Zhang}}, \bibinfo {author}
  {\bibfnamefont {H.-K.}\ \bibnamefont {Lo}},\ and\ \bibinfo {author}
  {\bibfnamefont {J.-W.}\ \bibnamefont {Pan}},\ }\bibfield  {title} {\bibinfo
  {title} {Secure quantum key distribution with realistic devices},\
  }\href@noop {} {\bibfield  {journal} {\bibinfo  {journal} {Reviews of Modern
  Physics}\ }\textbf {\bibinfo {volume} {92}},\ \bibinfo {pages} {025002}
  (\bibinfo {year} {2020})}\BibitemShut {NoStop}%
\bibitem [{\citenamefont {Pirandola}\ \emph {et~al.}(2020)\citenamefont
  {Pirandola}, \citenamefont {Andersen}, \citenamefont {Banchi}, \citenamefont
  {Berta}, \citenamefont {Bunandar}, \citenamefont {Colbeck}, \citenamefont
  {Englund}, \citenamefont {Gehring}, \citenamefont {Lupo}, \citenamefont
  {Ottaviani}, \citenamefont {Pereira}, \citenamefont {Razavi}, \citenamefont
  {Shaari}, \citenamefont {Tomamichel}, \citenamefont {Usenko}, \citenamefont
  {Vallone}, \citenamefont {Villoresi},\ and\ \citenamefont
  {Wallden}}]{pirandola2020advances}%
  \BibitemOpen
  \bibfield  {author} {\bibinfo {author} {\bibfnamefont {S.}~\bibnamefont
  {Pirandola}}, \bibinfo {author} {\bibfnamefont {U.~L.}\ \bibnamefont
  {Andersen}}, \bibinfo {author} {\bibfnamefont {L.}~\bibnamefont {Banchi}},
  \bibinfo {author} {\bibfnamefont {M.}~\bibnamefont {Berta}}, \bibinfo
  {author} {\bibfnamefont {D.}~\bibnamefont {Bunandar}}, \bibinfo {author}
  {\bibfnamefont {R.}~\bibnamefont {Colbeck}}, \bibinfo {author} {\bibfnamefont
  {D.}~\bibnamefont {Englund}}, \bibinfo {author} {\bibfnamefont
  {T.}~\bibnamefont {Gehring}}, \bibinfo {author} {\bibfnamefont
  {C.}~\bibnamefont {Lupo}}, \bibinfo {author} {\bibfnamefont {C.}~\bibnamefont
  {Ottaviani}}, \bibinfo {author} {\bibfnamefont {J.}~\bibnamefont {Pereira}},
  \bibinfo {author} {\bibfnamefont {M.}~\bibnamefont {Razavi}}, \bibinfo
  {author} {\bibfnamefont {J.~S.}\ \bibnamefont {Shaari}}, \bibinfo {author}
  {\bibfnamefont {M.}~\bibnamefont {Tomamichel}}, \bibinfo {author}
  {\bibfnamefont {V.~C.}\ \bibnamefont {Usenko}}, \bibinfo {author}
  {\bibfnamefont {G.}~\bibnamefont {Vallone}}, \bibinfo {author} {\bibfnamefont
  {P.}~\bibnamefont {Villoresi}},\ and\ \bibinfo {author} {\bibfnamefont
  {P.}~\bibnamefont {Wallden}},\ }\bibfield  {title} {\bibinfo {title}
  {Advances in quantum cryptography},\ }\href@noop {} {\bibfield  {journal}
  {\bibinfo  {journal} {Advances in Optics and Photonics}\ }\textbf {\bibinfo
  {volume} {12}},\ \bibinfo {pages} {1012} (\bibinfo {year}
  {2020})}\BibitemShut {NoStop}%
\bibitem [{\citenamefont {Scarani}\ \emph {et~al.}(2009)\citenamefont
  {Scarani}, \citenamefont {Bechmann-Pasquinucci}, \citenamefont {Cerf},
  \citenamefont {Du{\v{s}}ek}, \citenamefont {L{\"u}tkenhaus},\ and\
  \citenamefont {Peev}}]{scarani2009security}%
  \BibitemOpen
  \bibfield  {author} {\bibinfo {author} {\bibfnamefont {V.}~\bibnamefont
  {Scarani}}, \bibinfo {author} {\bibfnamefont {H.}~\bibnamefont
  {Bechmann-Pasquinucci}}, \bibinfo {author} {\bibfnamefont {N.~J.}\
  \bibnamefont {Cerf}}, \bibinfo {author} {\bibfnamefont {M.}~\bibnamefont
  {Du{\v{s}}ek}}, \bibinfo {author} {\bibfnamefont {N.}~\bibnamefont
  {L{\"u}tkenhaus}},\ and\ \bibinfo {author} {\bibfnamefont {M.}~\bibnamefont
  {Peev}},\ }\bibfield  {title} {\bibinfo {title} {The security of practical
  quantum key distribution},\ }\href@noop {} {\bibfield  {journal} {\bibinfo
  {journal} {Reviews of Modern Physics}\ }\textbf {\bibinfo {volume} {81}},\
  \bibinfo {pages} {1301} (\bibinfo {year} {2009})}\BibitemShut {NoStop}%
\bibitem [{\citenamefont {Liu}\ \emph {et~al.}(2023)\citenamefont {Liu},
  \citenamefont {Zhang}, \citenamefont {Jiang}, \citenamefont {Chen},
  \citenamefont {Zhang}, \citenamefont {Pan}, \citenamefont {Ma}, \citenamefont
  {Dong}, \citenamefont {Xiong}, \citenamefont {Zhang} \emph
  {et~al.}}]{liu2023experimental}%
  \BibitemOpen
  \bibfield  {author} {\bibinfo {author} {\bibfnamefont {Y.}~\bibnamefont
  {Liu}}, \bibinfo {author} {\bibfnamefont {W.-J.}\ \bibnamefont {Zhang}},
  \bibinfo {author} {\bibfnamefont {C.}~\bibnamefont {Jiang}}, \bibinfo
  {author} {\bibfnamefont {J.-P.}\ \bibnamefont {Chen}}, \bibinfo {author}
  {\bibfnamefont {C.}~\bibnamefont {Zhang}}, \bibinfo {author} {\bibfnamefont
  {W.-X.}\ \bibnamefont {Pan}}, \bibinfo {author} {\bibfnamefont
  {D.}~\bibnamefont {Ma}}, \bibinfo {author} {\bibfnamefont {H.}~\bibnamefont
  {Dong}}, \bibinfo {author} {\bibfnamefont {J.-M.}\ \bibnamefont {Xiong}},
  \bibinfo {author} {\bibfnamefont {C.-J.}\ \bibnamefont {Zhang}}, \emph
  {et~al.},\ }\bibfield  {title} {\bibinfo {title} {Experimental twin-field
  quantum key distribution over 1000 km fiber distance},\ }\href@noop {}
  {\bibfield  {journal} {\bibinfo  {journal} {Physical Review Letters}\
  }\textbf {\bibinfo {volume} {130}},\ \bibinfo {pages} {210801} (\bibinfo
  {year} {2023})}\BibitemShut {NoStop}%
\bibitem [{\citenamefont {Pittaluga}\ \emph {et~al.}(2025)\citenamefont
  {Pittaluga}, \citenamefont {Lo}, \citenamefont {Brzosko}, \citenamefont
  {Woodward}, \citenamefont {Scalcon}, \citenamefont {Winnel}, \citenamefont
  {Roger}, \citenamefont {Dynes}, \citenamefont {Owen}, \citenamefont
  {Ju{\'a}rez} \emph {et~al.}}]{pittaluga2025long}%
  \BibitemOpen
  \bibfield  {author} {\bibinfo {author} {\bibfnamefont {M.}~\bibnamefont
  {Pittaluga}}, \bibinfo {author} {\bibfnamefont {Y.~S.}\ \bibnamefont {Lo}},
  \bibinfo {author} {\bibfnamefont {A.}~\bibnamefont {Brzosko}}, \bibinfo
  {author} {\bibfnamefont {R.~I.}\ \bibnamefont {Woodward}}, \bibinfo {author}
  {\bibfnamefont {D.}~\bibnamefont {Scalcon}}, \bibinfo {author} {\bibfnamefont
  {M.~S.}\ \bibnamefont {Winnel}}, \bibinfo {author} {\bibfnamefont
  {T.}~\bibnamefont {Roger}}, \bibinfo {author} {\bibfnamefont {J.~F.}\
  \bibnamefont {Dynes}}, \bibinfo {author} {\bibfnamefont {K.~A.}\ \bibnamefont
  {Owen}}, \bibinfo {author} {\bibfnamefont {S.}~\bibnamefont {Ju{\'a}rez}},
  \emph {et~al.},\ }\bibfield  {title} {\bibinfo {title} {Long-distance
  coherent quantum communications in deployed telecom networks},\ }\href@noop
  {} {\bibfield  {journal} {\bibinfo  {journal} {Nature}\ }\textbf {\bibinfo
  {volume} {640}},\ \bibinfo {pages} {911} (\bibinfo {year}
  {2025})}\BibitemShut {NoStop}%
\bibitem [{\citenamefont {Pittaluga}\ \emph {et~al.}(2021)\citenamefont
  {Pittaluga}, \citenamefont {Minder}, \citenamefont {Lucamarini},
  \citenamefont {Sanzaro}, \citenamefont {Woodward}, \citenamefont {Li},
  \citenamefont {Yuan},\ and\ \citenamefont {Shields}}]{pittaluga2021600}%
  \BibitemOpen
  \bibfield  {author} {\bibinfo {author} {\bibfnamefont {M.}~\bibnamefont
  {Pittaluga}}, \bibinfo {author} {\bibfnamefont {M.}~\bibnamefont {Minder}},
  \bibinfo {author} {\bibfnamefont {M.}~\bibnamefont {Lucamarini}}, \bibinfo
  {author} {\bibfnamefont {M.}~\bibnamefont {Sanzaro}}, \bibinfo {author}
  {\bibfnamefont {R.~I.}\ \bibnamefont {Woodward}}, \bibinfo {author}
  {\bibfnamefont {M.-J.}\ \bibnamefont {Li}}, \bibinfo {author} {\bibfnamefont
  {Z.}~\bibnamefont {Yuan}},\ and\ \bibinfo {author} {\bibfnamefont {A.~J.}\
  \bibnamefont {Shields}},\ }\bibfield  {title} {\bibinfo {title} {600-km
  repeater-like quantum communications with dual-band stabilization},\
  }\href@noop {} {\bibfield  {journal} {\bibinfo  {journal} {Nature Photonics}\
  }\textbf {\bibinfo {volume} {15}},\ \bibinfo {pages} {530} (\bibinfo {year}
  {2021})}\BibitemShut {NoStop}%
\bibitem [{\citenamefont {Chen}\ \emph {et~al.}(2021)\citenamefont {Chen},
  \citenamefont {Zhang}, \citenamefont {Liu}, \citenamefont {Jiang},
  \citenamefont {Zhang}, \citenamefont {Han}, \citenamefont {Ma}, \citenamefont
  {Hu}, \citenamefont {Li}, \citenamefont {Liu}, \citenamefont {Zhou},
  \citenamefont {Jiang}, \citenamefont {Chen}, \citenamefont {Li},
  \citenamefont {You}, \citenamefont {Wang}, \citenamefont {Wang},
  \citenamefont {Zhang},\ and\ \citenamefont {Pan}}]{chen2021twin}%
  \BibitemOpen
  \bibfield  {author} {\bibinfo {author} {\bibfnamefont {J.-P.}\ \bibnamefont
  {Chen}}, \bibinfo {author} {\bibfnamefont {C.}~\bibnamefont {Zhang}},
  \bibinfo {author} {\bibfnamefont {Y.}~\bibnamefont {Liu}}, \bibinfo {author}
  {\bibfnamefont {C.}~\bibnamefont {Jiang}}, \bibinfo {author} {\bibfnamefont
  {W.-J.}\ \bibnamefont {Zhang}}, \bibinfo {author} {\bibfnamefont {Z.-Y.}\
  \bibnamefont {Han}}, \bibinfo {author} {\bibfnamefont {S.-Z.}\ \bibnamefont
  {Ma}}, \bibinfo {author} {\bibfnamefont {X.-L.}\ \bibnamefont {Hu}}, \bibinfo
  {author} {\bibfnamefont {Y.-H.}\ \bibnamefont {Li}}, \bibinfo {author}
  {\bibfnamefont {H.}~\bibnamefont {Liu}}, \bibinfo {author} {\bibfnamefont
  {F.}~\bibnamefont {Zhou}}, \bibinfo {author} {\bibfnamefont {H.-F.}\
  \bibnamefont {Jiang}}, \bibinfo {author} {\bibfnamefont {T.-Y.}\ \bibnamefont
  {Chen}}, \bibinfo {author} {\bibfnamefont {H.}~\bibnamefont {Li}}, \bibinfo
  {author} {\bibfnamefont {L.-X.}\ \bibnamefont {You}}, \bibinfo {author}
  {\bibfnamefont {Z.}~\bibnamefont {Wang}}, \bibinfo {author} {\bibfnamefont
  {X.-B.}\ \bibnamefont {Wang}}, \bibinfo {author} {\bibfnamefont
  {Q.}~\bibnamefont {Zhang}},\ and\ \bibinfo {author} {\bibfnamefont {J.-W.}\
  \bibnamefont {Pan}},\ }\bibfield  {title} {\bibinfo {title} {Twin-field
  quantum key distribution over 511 km optical fiber linking two distant
  metropolitans},\ }\href@noop {} {\bibfield  {journal} {\bibinfo  {journal}
  {Nature Photonics}\ }\textbf {\bibinfo {volume} {15}},\ \bibinfo {pages}
  {570} (\bibinfo {year} {2021})}\BibitemShut {NoStop}%
\bibitem [{\citenamefont {Liu}\ \emph {et~al.}(2021)\citenamefont {Liu},
  \citenamefont {Jiang}, \citenamefont {Zhu}, \citenamefont {Zou},
  \citenamefont {Yu}, \citenamefont {Hu}, \citenamefont {Xu}, \citenamefont
  {Ma}, \citenamefont {Han}, \citenamefont {Chen}, \citenamefont {Dai},
  \citenamefont {Tang}, \citenamefont {Zhang}, \citenamefont {Li},
  \citenamefont {You}, \citenamefont {Wang}, \citenamefont {Hua}, \citenamefont
  {Hu}, \citenamefont {Zhang}, \citenamefont {Zhou}, \citenamefont {Zhang},
  \citenamefont {Wang}, \citenamefont {Chen},\ and\ \citenamefont
  {Pan}}]{liu2021field}%
  \BibitemOpen
  \bibfield  {author} {\bibinfo {author} {\bibfnamefont {H.}~\bibnamefont
  {Liu}}, \bibinfo {author} {\bibfnamefont {C.}~\bibnamefont {Jiang}}, \bibinfo
  {author} {\bibfnamefont {H.-T.}\ \bibnamefont {Zhu}}, \bibinfo {author}
  {\bibfnamefont {M.}~\bibnamefont {Zou}}, \bibinfo {author} {\bibfnamefont
  {Z.-W.}\ \bibnamefont {Yu}}, \bibinfo {author} {\bibfnamefont {X.-L.}\
  \bibnamefont {Hu}}, \bibinfo {author} {\bibfnamefont {H.}~\bibnamefont {Xu}},
  \bibinfo {author} {\bibfnamefont {S.}~\bibnamefont {Ma}}, \bibinfo {author}
  {\bibfnamefont {Z.}~\bibnamefont {Han}}, \bibinfo {author} {\bibfnamefont
  {J.-P.}\ \bibnamefont {Chen}}, \bibinfo {author} {\bibfnamefont
  {Y.}~\bibnamefont {Dai}}, \bibinfo {author} {\bibfnamefont {S.-B.}\
  \bibnamefont {Tang}}, \bibinfo {author} {\bibfnamefont {W.}~\bibnamefont
  {Zhang}}, \bibinfo {author} {\bibfnamefont {H.}~\bibnamefont {Li}}, \bibinfo
  {author} {\bibfnamefont {L.}~\bibnamefont {You}}, \bibinfo {author}
  {\bibfnamefont {Z.}~\bibnamefont {Wang}}, \bibinfo {author} {\bibfnamefont
  {Y.}~\bibnamefont {Hua}}, \bibinfo {author} {\bibfnamefont {H.}~\bibnamefont
  {Hu}}, \bibinfo {author} {\bibfnamefont {H.}~\bibnamefont {Zhang}}, \bibinfo
  {author} {\bibfnamefont {F.}~\bibnamefont {Zhou}}, \bibinfo {author}
  {\bibfnamefont {Q.}~\bibnamefont {Zhang}}, \bibinfo {author} {\bibfnamefont
  {X.-B.}\ \bibnamefont {Wang}}, \bibinfo {author} {\bibfnamefont {T.-Y.}\
  \bibnamefont {Chen}},\ and\ \bibinfo {author} {\bibfnamefont {J.-W.}\
  \bibnamefont {Pan}},\ }\bibfield  {title} {\bibinfo {title} {Field test of
  twin-field quantum key distribution through sending-or-not-sending over 428
  km},\ }\href@noop {} {\bibfield  {journal} {\bibinfo  {journal} {Physical
  Review Letters}\ }\textbf {\bibinfo {volume} {126}},\ \bibinfo {pages}
  {250502} (\bibinfo {year} {2021})}\BibitemShut {NoStop}%
\bibitem [{\citenamefont {Wang}\ \emph {et~al.}(2022)\citenamefont {Wang},
  \citenamefont {Yin}, \citenamefont {He}, \citenamefont {Chen}, \citenamefont
  {Wang}, \citenamefont {Ye}, \citenamefont {Zhou}, \citenamefont {Fan-Yuan},
  \citenamefont {Wang}, \citenamefont {Chen}, \citenamefont {Zhu},
  \citenamefont {Morozov}, \citenamefont {Divochiy}, \citenamefont {Zhou},
  \citenamefont {Guo},\ and\ \citenamefont {Han}}]{wang2022twin}%
  \BibitemOpen
  \bibfield  {author} {\bibinfo {author} {\bibfnamefont {S.}~\bibnamefont
  {Wang}}, \bibinfo {author} {\bibfnamefont {Z.-Q.}\ \bibnamefont {Yin}},
  \bibinfo {author} {\bibfnamefont {D.-Y.}\ \bibnamefont {He}}, \bibinfo
  {author} {\bibfnamefont {W.}~\bibnamefont {Chen}}, \bibinfo {author}
  {\bibfnamefont {R.-Q.}\ \bibnamefont {Wang}}, \bibinfo {author}
  {\bibfnamefont {P.}~\bibnamefont {Ye}}, \bibinfo {author} {\bibfnamefont
  {Y.}~\bibnamefont {Zhou}}, \bibinfo {author} {\bibfnamefont {G.-J.}\
  \bibnamefont {Fan-Yuan}}, \bibinfo {author} {\bibfnamefont {F.-X.}\
  \bibnamefont {Wang}}, \bibinfo {author} {\bibfnamefont {W.}~\bibnamefont
  {Chen}}, \bibinfo {author} {\bibfnamefont {Y.-G.}\ \bibnamefont {Zhu}},
  \bibinfo {author} {\bibfnamefont {P.~V.}\ \bibnamefont {Morozov}}, \bibinfo
  {author} {\bibfnamefont {A.~V.}\ \bibnamefont {Divochiy}}, \bibinfo {author}
  {\bibfnamefont {Z.}~\bibnamefont {Zhou}}, \bibinfo {author} {\bibfnamefont
  {G.-C.}\ \bibnamefont {Guo}},\ and\ \bibinfo {author} {\bibfnamefont {Z.-F.}\
  \bibnamefont {Han}},\ }\bibfield  {title} {\bibinfo {title} {Twin-field
  quantum key distribution over 830-km fibre},\ }\href@noop {} {\bibfield
  {journal} {\bibinfo  {journal} {Nature Photonics}\ }\textbf {\bibinfo
  {volume} {16}},\ \bibinfo {pages} {154} (\bibinfo {year} {2022})}\BibitemShut
  {NoStop}%
\bibitem [{\citenamefont {Zhou}\ \emph {et~al.}(2023)\citenamefont {Zhou},
  \citenamefont {Lin}, \citenamefont {Jing},\ and\ \citenamefont
  {Yuan}}]{zhou2023twin}%
  \BibitemOpen
  \bibfield  {author} {\bibinfo {author} {\bibfnamefont {L.}~\bibnamefont
  {Zhou}}, \bibinfo {author} {\bibfnamefont {J.}~\bibnamefont {Lin}}, \bibinfo
  {author} {\bibfnamefont {Y.}~\bibnamefont {Jing}},\ and\ \bibinfo {author}
  {\bibfnamefont {Z.}~\bibnamefont {Yuan}},\ }\bibfield  {title} {\bibinfo
  {title} {Twin-field quantum key distribution without optical frequency
  dissemination},\ }\href@noop {} {\bibfield  {journal} {\bibinfo  {journal}
  {nature communications}\ }\textbf {\bibinfo {volume} {14}},\ \bibinfo {pages}
  {928} (\bibinfo {year} {2023})}\BibitemShut {NoStop}%
\bibitem [{\citenamefont {Li}\ \emph {et~al.}(2023)\citenamefont {Li},
  \citenamefont {Zhang}, \citenamefont {Tan}, \citenamefont {Lu}, \citenamefont
  {Liao}, \citenamefont {Huang}, \citenamefont {Li}, \citenamefont {Wang},
  \citenamefont {Mao}, \citenamefont {Yan} \emph {et~al.}}]{li2023high}%
  \BibitemOpen
  \bibfield  {author} {\bibinfo {author} {\bibfnamefont {W.}~\bibnamefont
  {Li}}, \bibinfo {author} {\bibfnamefont {L.}~\bibnamefont {Zhang}}, \bibinfo
  {author} {\bibfnamefont {H.}~\bibnamefont {Tan}}, \bibinfo {author}
  {\bibfnamefont {Y.}~\bibnamefont {Lu}}, \bibinfo {author} {\bibfnamefont
  {S.-K.}\ \bibnamefont {Liao}}, \bibinfo {author} {\bibfnamefont
  {J.}~\bibnamefont {Huang}}, \bibinfo {author} {\bibfnamefont
  {H.}~\bibnamefont {Li}}, \bibinfo {author} {\bibfnamefont {Z.}~\bibnamefont
  {Wang}}, \bibinfo {author} {\bibfnamefont {H.-K.}\ \bibnamefont {Mao}},
  \bibinfo {author} {\bibfnamefont {B.}~\bibnamefont {Yan}}, \emph {et~al.},\
  }\bibfield  {title} {\bibinfo {title} {High-rate quantum key distribution
  exceeding 110 mb s--1},\ }\href@noop {} {\bibfield  {journal} {\bibinfo
  {journal} {Nature photonics}\ }\textbf {\bibinfo {volume} {17}},\ \bibinfo
  {pages} {416} (\bibinfo {year} {2023})}\BibitemShut {NoStop}%
\bibitem [{\citenamefont {Lo}\ \emph {et~al.}(2012)\citenamefont {Lo},
  \citenamefont {Curty},\ and\ \citenamefont {Qi}}]{lo2012measurement}%
  \BibitemOpen
  \bibfield  {author} {\bibinfo {author} {\bibfnamefont {H.-K.}\ \bibnamefont
  {Lo}}, \bibinfo {author} {\bibfnamefont {M.}~\bibnamefont {Curty}},\ and\
  \bibinfo {author} {\bibfnamefont {B.}~\bibnamefont {Qi}},\ }\bibfield
  {title} {\bibinfo {title} {Measurement-device-independent quantum key
  distribution},\ }\href@noop {} {\bibfield  {journal} {\bibinfo  {journal}
  {Physical Review Letters}\ }\textbf {\bibinfo {volume} {108}},\ \bibinfo
  {pages} {130503} (\bibinfo {year} {2012})}\BibitemShut {NoStop}%
\bibitem [{\citenamefont {Braunstein}\ and\ \citenamefont
  {Pirandola}(2012)}]{braunstein2012side}%
  \BibitemOpen
  \bibfield  {author} {\bibinfo {author} {\bibfnamefont {S.~L.}\ \bibnamefont
  {Braunstein}}\ and\ \bibinfo {author} {\bibfnamefont {S.}~\bibnamefont
  {Pirandola}},\ }\bibfield  {title} {\bibinfo {title} {Side-channel-free
  quantum key distribution},\ }\href@noop {} {\bibfield  {journal} {\bibinfo
  {journal} {Physical Review Letters}\ }\textbf {\bibinfo {volume} {108}},\
  \bibinfo {pages} {130502} (\bibinfo {year} {2012})}\BibitemShut {NoStop}%
\bibitem [{\citenamefont {Wang}(2013)}]{wang2013three}%
  \BibitemOpen
  \bibfield  {author} {\bibinfo {author} {\bibfnamefont {X.-B.}\ \bibnamefont
  {Wang}},\ }\bibfield  {title} {\bibinfo {title} {Three-intensity decoy-state
  method for device-independent quantum key distribution with basis-dependent
  errors},\ }\href@noop {} {\bibfield  {journal} {\bibinfo  {journal} {Physical
  Review A}\ }\textbf {\bibinfo {volume} {87}},\ \bibinfo {pages} {012320}
  (\bibinfo {year} {2013})}\BibitemShut {NoStop}%
\bibitem [{\citenamefont {Zhou}\ \emph {et~al.}(2016)\citenamefont {Zhou},
  \citenamefont {Yu},\ and\ \citenamefont {Wang}}]{zhou2016making}%
  \BibitemOpen
  \bibfield  {author} {\bibinfo {author} {\bibfnamefont {Y.-H.}\ \bibnamefont
  {Zhou}}, \bibinfo {author} {\bibfnamefont {Z.-W.}\ \bibnamefont {Yu}},\ and\
  \bibinfo {author} {\bibfnamefont {X.-B.}\ \bibnamefont {Wang}},\ }\bibfield
  {title} {\bibinfo {title} {Making the decoy-state
  measurement-device-independent quantum key distribution practically useful},\
  }\href@noop {} {\bibfield  {journal} {\bibinfo  {journal} {Physical Review
  A}\ }\textbf {\bibinfo {volume} {93}},\ \bibinfo {pages} {042324} (\bibinfo
  {year} {2016})}\BibitemShut {NoStop}%
\bibitem [{\citenamefont {Jiang}\ \emph {et~al.}(2021)\citenamefont {Jiang},
  \citenamefont {Yu}, \citenamefont {Hu},\ and\ \citenamefont
  {Wang}}]{jiang2021higher}%
  \BibitemOpen
  \bibfield  {author} {\bibinfo {author} {\bibfnamefont {C.}~\bibnamefont
  {Jiang}}, \bibinfo {author} {\bibfnamefont {Z.-W.}\ \bibnamefont {Yu}},
  \bibinfo {author} {\bibfnamefont {X.-L.}\ \bibnamefont {Hu}},\ and\ \bibinfo
  {author} {\bibfnamefont {X.-B.}\ \bibnamefont {Wang}},\ }\bibfield  {title}
  {\bibinfo {title} {Higher key rate of measurement-device-independent quantum
  key distribution through joint data processing},\ }\href@noop {} {\bibfield
  {journal} {\bibinfo  {journal} {Physical Review A}\ }\textbf {\bibinfo
  {volume} {103}},\ \bibinfo {pages} {012402} (\bibinfo {year}
  {2021})}\BibitemShut {NoStop}%
\bibitem [{\citenamefont {Lucamarini}\ \emph {et~al.}(2018)\citenamefont
  {Lucamarini}, \citenamefont {Yuan}, \citenamefont {Dynes},\ and\
  \citenamefont {Shields}}]{lu2018overcoming}%
  \BibitemOpen
  \bibfield  {author} {\bibinfo {author} {\bibfnamefont {M.}~\bibnamefont
  {Lucamarini}}, \bibinfo {author} {\bibfnamefont {Z.~L.}\ \bibnamefont
  {Yuan}}, \bibinfo {author} {\bibfnamefont {J.~F.}\ \bibnamefont {Dynes}},\
  and\ \bibinfo {author} {\bibfnamefont {A.~J.}\ \bibnamefont {Shields}},\
  }\bibfield  {title} {\bibinfo {title} {Overcoming the rate--distance limit of
  quantum key distribution without quantum repeaters},\ }\href@noop {}
  {\bibfield  {journal} {\bibinfo  {journal} {Nature}\ }\textbf {\bibinfo
  {volume} {557}},\ \bibinfo {pages} {400} (\bibinfo {year}
  {2018})}\BibitemShut {NoStop}%
\bibitem [{\citenamefont {Wang}\ \emph {et~al.}(2018)\citenamefont {Wang},
  \citenamefont {Yu},\ and\ \citenamefont {Hu}}]{wang2018twin}%
  \BibitemOpen
  \bibfield  {author} {\bibinfo {author} {\bibfnamefont {X.-B.}\ \bibnamefont
  {Wang}}, \bibinfo {author} {\bibfnamefont {Z.-W.}\ \bibnamefont {Yu}},\ and\
  \bibinfo {author} {\bibfnamefont {X.-L.}\ \bibnamefont {Hu}},\ }\bibfield
  {title} {\bibinfo {title} {Twin-field quantum key distribution with large
  misalignment error},\ }\href@noop {} {\bibfield  {journal} {\bibinfo
  {journal} {Physical Review A}\ }\textbf {\bibinfo {volume} {98}},\ \bibinfo
  {pages} {062323} (\bibinfo {year} {2018})}\BibitemShut {NoStop}%
\bibitem [{\citenamefont {Ma}\ \emph {et~al.}(2018)\citenamefont {Ma},
  \citenamefont {Zeng},\ and\ \citenamefont {Zhou}}]{ma2018phase}%
  \BibitemOpen
  \bibfield  {author} {\bibinfo {author} {\bibfnamefont {X.}~\bibnamefont
  {Ma}}, \bibinfo {author} {\bibfnamefont {P.}~\bibnamefont {Zeng}},\ and\
  \bibinfo {author} {\bibfnamefont {H.}~\bibnamefont {Zhou}},\ }\bibfield
  {title} {\bibinfo {title} {Phase-matching quantum key distribution},\
  }\href@noop {} {\bibfield  {journal} {\bibinfo  {journal} {Physical Review
  X}\ }\textbf {\bibinfo {volume} {8}},\ \bibinfo {pages} {031043} (\bibinfo
  {year} {2018})}\BibitemShut {NoStop}%
\bibitem [{\citenamefont {Lin}\ and\ \citenamefont
  {L{\"u}tkenhaus}(2018)}]{lin2018simple}%
  \BibitemOpen
  \bibfield  {author} {\bibinfo {author} {\bibfnamefont {J.}~\bibnamefont
  {Lin}}\ and\ \bibinfo {author} {\bibfnamefont {N.}~\bibnamefont
  {L{\"u}tkenhaus}},\ }\bibfield  {title} {\bibinfo {title} {Simple security
  analysis of phase-matching measurement-device-independent quantum key
  distribution},\ }\href@noop {} {\bibfield  {journal} {\bibinfo  {journal}
  {Physical Review A}\ }\textbf {\bibinfo {volume} {98}},\ \bibinfo {pages}
  {042332} (\bibinfo {year} {2018})}\BibitemShut {NoStop}%
\bibitem [{\citenamefont {Curty}\ \emph {et~al.}(2019)\citenamefont {Curty},
  \citenamefont {Azuma},\ and\ \citenamefont {Lo}}]{curty2018simple}%
  \BibitemOpen
  \bibfield  {author} {\bibinfo {author} {\bibfnamefont {M.}~\bibnamefont
  {Curty}}, \bibinfo {author} {\bibfnamefont {K.}~\bibnamefont {Azuma}},\ and\
  \bibinfo {author} {\bibfnamefont {H.-K.}\ \bibnamefont {Lo}},\ }\bibfield
  {title} {\bibinfo {title} {Simple security proof of twin-field type quantum
  key distribution protocol},\ }\href@noop {} {\bibfield  {journal} {\bibinfo
  {journal} {NPJ Quantum Information}\ }\textbf {\bibinfo {volume} {5}},\
  \bibinfo {pages} {64} (\bibinfo {year} {2019})}\BibitemShut {NoStop}%
\bibitem [{\citenamefont {Cui}\ \emph {et~al.}(2019)\citenamefont {Cui},
  \citenamefont {Yin}, \citenamefont {Wang}, \citenamefont {Chen},
  \citenamefont {Wang}, \citenamefont {Guo},\ and\ \citenamefont
  {Han}}]{cui2019twin}%
  \BibitemOpen
  \bibfield  {author} {\bibinfo {author} {\bibfnamefont {C.}~\bibnamefont
  {Cui}}, \bibinfo {author} {\bibfnamefont {Z.-Q.}\ \bibnamefont {Yin}},
  \bibinfo {author} {\bibfnamefont {R.}~\bibnamefont {Wang}}, \bibinfo {author}
  {\bibfnamefont {W.}~\bibnamefont {Chen}}, \bibinfo {author} {\bibfnamefont
  {S.}~\bibnamefont {Wang}}, \bibinfo {author} {\bibfnamefont {G.-C.}\
  \bibnamefont {Guo}},\ and\ \bibinfo {author} {\bibfnamefont {Z.-F.}\
  \bibnamefont {Han}},\ }\bibfield  {title} {\bibinfo {title} {Twin-field
  quantum key distribution without phase postselection},\ }\href@noop {}
  {\bibfield  {journal} {\bibinfo  {journal} {Physical Review Applied}\
  }\textbf {\bibinfo {volume} {11}},\ \bibinfo {pages} {034053} (\bibinfo
  {year} {2019})}\BibitemShut {NoStop}%
\bibitem [{\citenamefont {Hwang}(2003)}]{hwang2003quantum}%
  \BibitemOpen
  \bibfield  {author} {\bibinfo {author} {\bibfnamefont {W.-Y.}\ \bibnamefont
  {Hwang}},\ }\bibfield  {title} {\bibinfo {title} {Quantum key distribution
  with high loss: toward global secure communication},\ }\href@noop {}
  {\bibfield  {journal} {\bibinfo  {journal} {Physical Review Letters}\
  }\textbf {\bibinfo {volume} {91}},\ \bibinfo {pages} {057901} (\bibinfo
  {year} {2003})}\BibitemShut {NoStop}%
\bibitem [{\citenamefont {Wang}(2005)}]{wang2005beating}%
  \BibitemOpen
  \bibfield  {author} {\bibinfo {author} {\bibfnamefont {X.-B.}\ \bibnamefont
  {Wang}},\ }\bibfield  {title} {\bibinfo {title} {Beating the
  photon-number-splitting attack in practical quantum cryptography},\
  }\href@noop {} {\bibfield  {journal} {\bibinfo  {journal} {Physical Review
  Letters}\ }\textbf {\bibinfo {volume} {94}},\ \bibinfo {pages} {230503}
  (\bibinfo {year} {2005})}\BibitemShut {NoStop}%
\bibitem [{\citenamefont {Lo}\ \emph {et~al.}(2005)\citenamefont {Lo},
  \citenamefont {Ma},\ and\ \citenamefont {Chen}}]{lo2005decoy}%
  \BibitemOpen
  \bibfield  {author} {\bibinfo {author} {\bibfnamefont {H.-K.}\ \bibnamefont
  {Lo}}, \bibinfo {author} {\bibfnamefont {X.}~\bibnamefont {Ma}},\ and\
  \bibinfo {author} {\bibfnamefont {K.}~\bibnamefont {Chen}},\ }\bibfield
  {title} {\bibinfo {title} {Decoy state quantum key distribution},\
  }\href@noop {} {\bibfield  {journal} {\bibinfo  {journal} {Physical Review
  Letters}\ }\textbf {\bibinfo {volume} {94}},\ \bibinfo {pages} {230504}
  (\bibinfo {year} {2005})}\BibitemShut {NoStop}%
\bibitem [{\citenamefont {Huang}\ \emph {et~al.}(2018)\citenamefont {Huang},
  \citenamefont {Sun}, \citenamefont {Liu},\ and\ \citenamefont
  {Makarov}}]{huang2018quantum}%
  \BibitemOpen
  \bibfield  {author} {\bibinfo {author} {\bibfnamefont {A.}~\bibnamefont
  {Huang}}, \bibinfo {author} {\bibfnamefont {S.-H.}\ \bibnamefont {Sun}},
  \bibinfo {author} {\bibfnamefont {Z.}~\bibnamefont {Liu}},\ and\ \bibinfo
  {author} {\bibfnamefont {V.}~\bibnamefont {Makarov}},\ }\bibfield  {title}
  {\bibinfo {title} {Quantum key distribution with distinguishable decoy
  states},\ }\href@noop {} {\bibfield  {journal} {\bibinfo  {journal} {Physical
  Review A}\ }\textbf {\bibinfo {volume} {98}},\ \bibinfo {pages} {012330}
  (\bibinfo {year} {2018})}\BibitemShut {NoStop}%
\bibitem [{\citenamefont {Gnanapandithan}\ \emph {et~al.}(2025)\citenamefont
  {Gnanapandithan}, \citenamefont {Qian},\ and\ \citenamefont
  {Lo}}]{gnanapandithan2025hidden}%
  \BibitemOpen
  \bibfield  {author} {\bibinfo {author} {\bibfnamefont {A.}~\bibnamefont
  {Gnanapandithan}}, \bibinfo {author} {\bibfnamefont {L.}~\bibnamefont
  {Qian}},\ and\ \bibinfo {author} {\bibfnamefont {H.-K.}\ \bibnamefont {Lo}},\
  }\bibfield  {title} {\bibinfo {title} {Hidden multidimensional modulation
  side channels in quantum protocols},\ }\href@noop {} {\bibfield  {journal}
  {\bibinfo  {journal} {Physical Review Letters}\ }\textbf {\bibinfo {volume}
  {134}},\ \bibinfo {pages} {130802} (\bibinfo {year} {2025})}\BibitemShut
  {NoStop}%
\bibitem [{\citenamefont {Wang}\ \emph {et~al.}(2019)\citenamefont {Wang},
  \citenamefont {Hu},\ and\ \citenamefont {Yu}}]{wang2019practical}%
  \BibitemOpen
  \bibfield  {author} {\bibinfo {author} {\bibfnamefont {X.-B.}\ \bibnamefont
  {Wang}}, \bibinfo {author} {\bibfnamefont {X.-L.}\ \bibnamefont {Hu}},\ and\
  \bibinfo {author} {\bibfnamefont {Z.-W.}\ \bibnamefont {Yu}},\ }\bibfield
  {title} {\bibinfo {title} {Practical long-distance side-channel-free quantum
  key distribution},\ }\href@noop {} {\bibfield  {journal} {\bibinfo  {journal}
  {Physical Review Applied}\ }\textbf {\bibinfo {volume} {12}},\ \bibinfo
  {pages} {054034} (\bibinfo {year} {2019})}\BibitemShut {NoStop}%
\bibitem [{\citenamefont {Zhang}\ \emph {et~al.}(2022)\citenamefont {Zhang},
  \citenamefont {Hu}, \citenamefont {Jiang}, \citenamefont {Chen},
  \citenamefont {Liu}, \citenamefont {Zhang}, \citenamefont {Yu}, \citenamefont
  {Li}, \citenamefont {You}, \citenamefont {Wang}, \citenamefont {Wang},
  \citenamefont {Zhang},\ and\ \citenamefont {Pan}}]{zhang2022experimental}%
  \BibitemOpen
  \bibfield  {author} {\bibinfo {author} {\bibfnamefont {C.}~\bibnamefont
  {Zhang}}, \bibinfo {author} {\bibfnamefont {X.-L.}\ \bibnamefont {Hu}},
  \bibinfo {author} {\bibfnamefont {C.}~\bibnamefont {Jiang}}, \bibinfo
  {author} {\bibfnamefont {J.-P.}\ \bibnamefont {Chen}}, \bibinfo {author}
  {\bibfnamefont {Y.}~\bibnamefont {Liu}}, \bibinfo {author} {\bibfnamefont
  {W.}~\bibnamefont {Zhang}}, \bibinfo {author} {\bibfnamefont {Z.-W.}\
  \bibnamefont {Yu}}, \bibinfo {author} {\bibfnamefont {H.}~\bibnamefont {Li}},
  \bibinfo {author} {\bibfnamefont {L.}~\bibnamefont {You}}, \bibinfo {author}
  {\bibfnamefont {Z.}~\bibnamefont {Wang}}, \bibinfo {author} {\bibfnamefont
  {X.-B.}\ \bibnamefont {Wang}}, \bibinfo {author} {\bibfnamefont
  {Q.}~\bibnamefont {Zhang}},\ and\ \bibinfo {author} {\bibfnamefont {J.-W.}\
  \bibnamefont {Pan}},\ }\bibfield  {title} {\bibinfo {title} {Experimental
  side-channel-secure quantum key distribution},\ }\href@noop {} {\bibfield
  {journal} {\bibinfo  {journal} {Physical Review Letter}\ }\textbf {\bibinfo
  {volume} {128}},\ \bibinfo {pages} {190503} (\bibinfo {year}
  {2022})}\BibitemShut {NoStop}%
\bibitem [{\citenamefont {Jiang}\ \emph {et~al.}(2023)\citenamefont {Jiang},
  \citenamefont {Yu}, \citenamefont {Hu},\ and\ \citenamefont
  {Wang}}]{jiang2023side}%
  \BibitemOpen
  \bibfield  {author} {\bibinfo {author} {\bibfnamefont {C.}~\bibnamefont
  {Jiang}}, \bibinfo {author} {\bibfnamefont {Z.-W.}\ \bibnamefont {Yu}},
  \bibinfo {author} {\bibfnamefont {X.-L.}\ \bibnamefont {Hu}},\ and\ \bibinfo
  {author} {\bibfnamefont {X.-B.}\ \bibnamefont {Wang}},\ }\bibfield  {title}
  {\bibinfo {title} {Side-channel-secure quantum key distribution with
  imperfect vacuum sources},\ }\href@noop {} {\bibfield  {journal} {\bibinfo
  {journal} {Physical Review Applied}\ }\textbf {\bibinfo {volume} {19}},\
  \bibinfo {pages} {064003} (\bibinfo {year} {2023})}\BibitemShut {NoStop}%
\bibitem [{\citenamefont {Jiang}\ \emph {et~al.}(2024)\citenamefont {Jiang},
  \citenamefont {Hu}, \citenamefont {Yu},\ and\ \citenamefont
  {Wang}}]{jiang2024side}%
  \BibitemOpen
  \bibfield  {author} {\bibinfo {author} {\bibfnamefont {C.}~\bibnamefont
  {Jiang}}, \bibinfo {author} {\bibfnamefont {X.-L.}\ \bibnamefont {Hu}},
  \bibinfo {author} {\bibfnamefont {Z.-W.}\ \bibnamefont {Yu}},\ and\ \bibinfo
  {author} {\bibfnamefont {X.-B.}\ \bibnamefont {Wang}},\ }\bibfield  {title}
  {\bibinfo {title} {Side-channel security of practical quantum key
  distribution},\ }\href@noop {} {\bibfield  {journal} {\bibinfo  {journal}
  {Physical Review Research}\ }\textbf {\bibinfo {volume} {6}},\ \bibinfo
  {pages} {013266} (\bibinfo {year} {2024})}\BibitemShut {NoStop}%
\bibitem [{\citenamefont {Gisin}\ \emph {et~al.}(2006)\citenamefont {Gisin},
  \citenamefont {Fasel}, \citenamefont {Kraus}, \citenamefont {Zbinden},\ and\
  \citenamefont {Ribordy}}]{gisin2006trojan}%
  \BibitemOpen
  \bibfield  {author} {\bibinfo {author} {\bibfnamefont {N.}~\bibnamefont
  {Gisin}}, \bibinfo {author} {\bibfnamefont {S.}~\bibnamefont {Fasel}},
  \bibinfo {author} {\bibfnamefont {B.}~\bibnamefont {Kraus}}, \bibinfo
  {author} {\bibfnamefont {H.}~\bibnamefont {Zbinden}},\ and\ \bibinfo {author}
  {\bibfnamefont {G.}~\bibnamefont {Ribordy}},\ }\bibfield  {title} {\bibinfo
  {title} {Trojan-horse attacks on quantum-key-distribution systems},\
  }\href@noop {} {\bibfield  {journal} {\bibinfo  {journal} {Physical Review
  A}\ }\textbf {\bibinfo {volume} {73}},\ \bibinfo {pages} {022320} (\bibinfo
  {year} {2006})}\BibitemShut {NoStop}%
\bibitem [{\citenamefont {Lucamarini}\ \emph {et~al.}(2015)\citenamefont
  {Lucamarini}, \citenamefont {Choi}, \citenamefont {Ward}, \citenamefont
  {Dynes}, \citenamefont {Yuan},\ and\ \citenamefont
  {Shields}}]{lucamarini2015practical}%
  \BibitemOpen
  \bibfield  {author} {\bibinfo {author} {\bibfnamefont {M.}~\bibnamefont
  {Lucamarini}}, \bibinfo {author} {\bibfnamefont {I.}~\bibnamefont {Choi}},
  \bibinfo {author} {\bibfnamefont {M.~B.}\ \bibnamefont {Ward}}, \bibinfo
  {author} {\bibfnamefont {J.~F.}\ \bibnamefont {Dynes}}, \bibinfo {author}
  {\bibfnamefont {Z.}~\bibnamefont {Yuan}},\ and\ \bibinfo {author}
  {\bibfnamefont {A.~J.}\ \bibnamefont {Shields}},\ }\bibfield  {title}
  {\bibinfo {title} {Practical security bounds against the trojan-horse attack
  in quantum key distribution},\ }\href@noop {} {\bibfield  {journal} {\bibinfo
   {journal} {Physical Review X}\ }\textbf {\bibinfo {volume} {5}},\ \bibinfo
  {pages} {031030} (\bibinfo {year} {2015})}\BibitemShut {NoStop}%
\bibitem [{\citenamefont {Tan}\ \emph {et~al.}(2021)\citenamefont {Tan},
  \citenamefont {Li}, \citenamefont {Zhang}, \citenamefont {Wei},\ and\
  \citenamefont {Xu}}]{tan2021chip}%
  \BibitemOpen
  \bibfield  {author} {\bibinfo {author} {\bibfnamefont {H.}~\bibnamefont
  {Tan}}, \bibinfo {author} {\bibfnamefont {W.}~\bibnamefont {Li}}, \bibinfo
  {author} {\bibfnamefont {L.}~\bibnamefont {Zhang}}, \bibinfo {author}
  {\bibfnamefont {K.}~\bibnamefont {Wei}},\ and\ \bibinfo {author}
  {\bibfnamefont {F.}~\bibnamefont {Xu}},\ }\bibfield  {title} {\bibinfo
  {title} {Chip-based quantum key distribution against trojan-horse attack},\
  }\href@noop {} {\bibfield  {journal} {\bibinfo  {journal} {Physical Review
  Applied}\ }\textbf {\bibinfo {volume} {15}},\ \bibinfo {pages} {064038}
  (\bibinfo {year} {2021})}\BibitemShut {NoStop}%
\bibitem [{\citenamefont {Roberts}\ \emph {et~al.}(2018)\citenamefont
  {Roberts}, \citenamefont {Pittaluga}, \citenamefont {Minder}, \citenamefont
  {Lucamarini}, \citenamefont {Dynes}, \citenamefont {Yuan},\ and\
  \citenamefont {Shields}}]{roberts2018patterning}%
  \BibitemOpen
  \bibfield  {author} {\bibinfo {author} {\bibfnamefont {G.}~\bibnamefont
  {Roberts}}, \bibinfo {author} {\bibfnamefont {M.}~\bibnamefont {Pittaluga}},
  \bibinfo {author} {\bibfnamefont {M.}~\bibnamefont {Minder}}, \bibinfo
  {author} {\bibfnamefont {M.}~\bibnamefont {Lucamarini}}, \bibinfo {author}
  {\bibfnamefont {J.}~\bibnamefont {Dynes}}, \bibinfo {author} {\bibfnamefont
  {Z.}~\bibnamefont {Yuan}},\ and\ \bibinfo {author} {\bibfnamefont
  {A.}~\bibnamefont {Shields}},\ }\bibfield  {title} {\bibinfo {title}
  {Patterning-effect mitigating intensity modulator for secure decoy-state
  quantum key distribution},\ }\href@noop {} {\bibfield  {journal} {\bibinfo
  {journal} {Optics letters}\ }\textbf {\bibinfo {volume} {43}},\ \bibinfo
  {pages} {5110} (\bibinfo {year} {2018})}\BibitemShut {NoStop}%
\bibitem [{\citenamefont {Zapatero}\ \emph {et~al.}(2021)\citenamefont
  {Zapatero}, \citenamefont {Navarrete}, \citenamefont {Tamaki},\ and\
  \citenamefont {Curty}}]{zapatero2021security}%
  \BibitemOpen
  \bibfield  {author} {\bibinfo {author} {\bibfnamefont {V.}~\bibnamefont
  {Zapatero}}, \bibinfo {author} {\bibfnamefont {{\'A}.}~\bibnamefont
  {Navarrete}}, \bibinfo {author} {\bibfnamefont {K.}~\bibnamefont {Tamaki}},\
  and\ \bibinfo {author} {\bibfnamefont {M.}~\bibnamefont {Curty}},\ }\bibfield
   {title} {\bibinfo {title} {Security of quantum key distribution with
  intensity correlations},\ }\href@noop {} {\bibfield  {journal} {\bibinfo
  {journal} {Quantum}\ }\textbf {\bibinfo {volume} {5}},\ \bibinfo {pages}
  {602} (\bibinfo {year} {2021})}\BibitemShut {NoStop}%
\bibitem [{\citenamefont {Pereira}\ \emph {et~al.}(2024)\citenamefont
  {Pereira}, \citenamefont {Curr{\'a}s-Lorenzo}, \citenamefont {Mizutani},
  \citenamefont {Rusca}, \citenamefont {Curty},\ and\ \citenamefont
  {Tamaki}}]{pereira2024quantum}%
  \BibitemOpen
  \bibfield  {author} {\bibinfo {author} {\bibfnamefont {M.}~\bibnamefont
  {Pereira}}, \bibinfo {author} {\bibfnamefont {G.}~\bibnamefont
  {Curr{\'a}s-Lorenzo}}, \bibinfo {author} {\bibfnamefont {A.}~\bibnamefont
  {Mizutani}}, \bibinfo {author} {\bibfnamefont {D.}~\bibnamefont {Rusca}},
  \bibinfo {author} {\bibfnamefont {M.}~\bibnamefont {Curty}},\ and\ \bibinfo
  {author} {\bibfnamefont {K.}~\bibnamefont {Tamaki}},\ }\bibfield  {title}
  {\bibinfo {title} {Quantum key distribution with unbounded pulse
  correlations},\ }\href@noop {} {\bibfield  {journal} {\bibinfo  {journal}
  {Quantum Science and Technology}\ }\textbf {\bibinfo {volume} {10}},\
  \bibinfo {pages} {015001} (\bibinfo {year} {2024})}\BibitemShut {NoStop}%
\bibitem [{\citenamefont {Chernoff}(1952)}]{chernoff1952measure}%
  \BibitemOpen
  \bibfield  {author} {\bibinfo {author} {\bibfnamefont {H.}~\bibnamefont
  {Chernoff}},\ }\bibfield  {title} {\bibinfo {title} {A measure of asymptotic
  efficiency for tests of a hypothesis based on the sum of observations},\
  }\href@noop {} {\bibfield  {journal} {\bibinfo  {journal} {The Annals of
  Mathematical Statistics}\ }\textbf {\bibinfo {volume} {23}},\ \bibinfo
  {pages} {493} (\bibinfo {year} {1952})}\BibitemShut {NoStop}%
\bibitem [{\citenamefont {Scarani}\ and\ \citenamefont
  {Renner}(2008)}]{scarani2008security}%
  \BibitemOpen
  \bibfield  {author} {\bibinfo {author} {\bibfnamefont {V.}~\bibnamefont
  {Scarani}}\ and\ \bibinfo {author} {\bibfnamefont {R.}~\bibnamefont
  {Renner}},\ }\bibfield  {title} {\bibinfo {title} {Security bounds for
  quantum cryptography with finite resources},\ }in\ \href@noop {} {\emph
  {\bibinfo {booktitle} {Workshop on Quantum Computation, Communication, and
  Cryptography}}}\ (\bibinfo {organization} {Springer},\ \bibinfo {year}
  {2008})\ pp.\ \bibinfo {pages} {83--95}\BibitemShut {NoStop}%
\bibitem [{\citenamefont {Sheridan}\ \emph {et~al.}(2010)\citenamefont
  {Sheridan}, \citenamefont {Le},\ and\ \citenamefont
  {Scarani}}]{sheridan2010finite}%
  \BibitemOpen
  \bibfield  {author} {\bibinfo {author} {\bibfnamefont {L.}~\bibnamefont
  {Sheridan}}, \bibinfo {author} {\bibfnamefont {T.~P.}\ \bibnamefont {Le}},\
  and\ \bibinfo {author} {\bibfnamefont {V.}~\bibnamefont {Scarani}},\
  }\bibfield  {title} {\bibinfo {title} {Finite-key security against coherent
  attacks in quantum key distribution},\ }\href@noop {} {\bibfield  {journal}
  {\bibinfo  {journal} {New Journal of Physics}\ }\textbf {\bibinfo {volume}
  {12}},\ \bibinfo {pages} {123019} (\bibinfo {year} {2010})}\BibitemShut
  {NoStop}%
\bibitem [{\citenamefont {Christandl}\ \emph {et~al.}(2009)\citenamefont
  {Christandl}, \citenamefont {K{\"o}nig},\ and\ \citenamefont
  {Renner}}]{christandl2009postselection}%
  \BibitemOpen
  \bibfield  {author} {\bibinfo {author} {\bibfnamefont {M.}~\bibnamefont
  {Christandl}}, \bibinfo {author} {\bibfnamefont {R.}~\bibnamefont
  {K{\"o}nig}},\ and\ \bibinfo {author} {\bibfnamefont {R.}~\bibnamefont
  {Renner}},\ }\bibfield  {title} {\bibinfo {title} {Postselection technique
  for quantum channels with applications to quantum cryptography},\ }\href@noop
  {} {\bibfield  {journal} {\bibinfo  {journal} {Physical Review Letters}\
  }\textbf {\bibinfo {volume} {102}},\ \bibinfo {pages} {020504} (\bibinfo
  {year} {2009})}\BibitemShut {NoStop}%
\bibitem [{\citenamefont {Jain}\ \emph
  {et~al.}(2014{\natexlab{a}})\citenamefont {Jain}, \citenamefont {Anisimova},
  \citenamefont {Khan}, \citenamefont {Makarov}, \citenamefont {Marquardt},\
  and\ \citenamefont {Leuchs}}]{jain2014trojan}%
  \BibitemOpen
  \bibfield  {author} {\bibinfo {author} {\bibfnamefont {N.}~\bibnamefont
  {Jain}}, \bibinfo {author} {\bibfnamefont {E.}~\bibnamefont {Anisimova}},
  \bibinfo {author} {\bibfnamefont {I.}~\bibnamefont {Khan}}, \bibinfo {author}
  {\bibfnamefont {V.}~\bibnamefont {Makarov}}, \bibinfo {author} {\bibfnamefont
  {C.}~\bibnamefont {Marquardt}},\ and\ \bibinfo {author} {\bibfnamefont
  {G.}~\bibnamefont {Leuchs}},\ }\bibfield  {title} {\bibinfo {title}
  {Trojan-horse attacks threaten the security of practical quantum
  cryptography},\ }\href@noop {} {\bibfield  {journal} {\bibinfo  {journal}
  {New Journal of Physics}\ }\textbf {\bibinfo {volume} {16}},\ \bibinfo
  {pages} {123030} (\bibinfo {year} {2014}{\natexlab{a}})}\BibitemShut
  {NoStop}%
\bibitem [{\citenamefont {Jain}\ \emph
  {et~al.}(2014{\natexlab{b}})\citenamefont {Jain}, \citenamefont {Stiller},
  \citenamefont {Khan}, \citenamefont {Makarov}, \citenamefont {Marquardt},\
  and\ \citenamefont {Leuchs}}]{jain2014risk}%
  \BibitemOpen
  \bibfield  {author} {\bibinfo {author} {\bibfnamefont {N.}~\bibnamefont
  {Jain}}, \bibinfo {author} {\bibfnamefont {B.}~\bibnamefont {Stiller}},
  \bibinfo {author} {\bibfnamefont {I.}~\bibnamefont {Khan}}, \bibinfo {author}
  {\bibfnamefont {V.}~\bibnamefont {Makarov}}, \bibinfo {author} {\bibfnamefont
  {C.}~\bibnamefont {Marquardt}},\ and\ \bibinfo {author} {\bibfnamefont
  {G.}~\bibnamefont {Leuchs}},\ }\bibfield  {title} {\bibinfo {title} {Risk
  analysis of trojan-horse attacks on practical quantum key distribution
  systems},\ }\href@noop {} {\bibfield  {journal} {\bibinfo  {journal} {IEEE
  Journal of Selected Topics in Quantum Electronics}\ }\textbf {\bibinfo
  {volume} {21}},\ \bibinfo {pages} {168} (\bibinfo {year}
  {2014}{\natexlab{b}})}\BibitemShut {NoStop}%
\bibitem [{\citenamefont {Ponosova}\ \emph {et~al.}(2022)\citenamefont
  {Ponosova}, \citenamefont {Ruzhitskaya}, \citenamefont {Chaiwongkhot},
  \citenamefont {Egorov}, \citenamefont {Makarov},\ and\ \citenamefont
  {Huang}}]{ponosova2022protecting}%
  \BibitemOpen
  \bibfield  {author} {\bibinfo {author} {\bibfnamefont {A.}~\bibnamefont
  {Ponosova}}, \bibinfo {author} {\bibfnamefont {D.}~\bibnamefont
  {Ruzhitskaya}}, \bibinfo {author} {\bibfnamefont {P.}~\bibnamefont
  {Chaiwongkhot}}, \bibinfo {author} {\bibfnamefont {V.}~\bibnamefont
  {Egorov}}, \bibinfo {author} {\bibfnamefont {V.}~\bibnamefont {Makarov}},\
  and\ \bibinfo {author} {\bibfnamefont {A.}~\bibnamefont {Huang}},\ }\bibfield
   {title} {\bibinfo {title} {Protecting fiber-optic quantum key distribution
  sources against light-injection attacks},\ }\href@noop {} {\bibfield
  {journal} {\bibinfo  {journal} {PRX Quantum}\ }\textbf {\bibinfo {volume}
  {3}},\ \bibinfo {pages} {040307} (\bibinfo {year} {2022})}\BibitemShut
  {NoStop}%
\end{thebibliography}%

\end{document}